\documentclass[12pt]{article}
\usepackage{amsmath}
\usepackage{latexsym}
\usepackage{float}
\usepackage{amssymb}
\usepackage{graphicx}
\begin{document}

\title{Polymer Brushes in Cylindrical Pores: Simulation versus
Scaling Theory}

\author{D. I. Dimitrov$^1$, A. Milchev$^{2,3}$, and K. Binder$^3$ \\
[\baselineskip]
$^{(1)}$ {\it Inorganic Chem. and  Phys. Chem. Dept, Univ. Food Technol.,} \\
{\it Maritza Blvd. 26, 4002 Plovdiv, Bulgaria }\\
$^{(2)}$ {\it Instute for Chemical Physics, Bulgarian Academy of Sciences,} \\
{\it 1113 Sofia, Bulgaria} \\
{\it$^{(3)}$ Institut f\"ur Physik, Johannes Gutenberg Universit\"at Mainz, }\\
{\it Staudinger Weg 7, 55099 Mainz, Germany}
}

\maketitle

\begin{abstract}
The structure of flexible polymers endgrafted in cylindrical
pores of diameter $D$ is studied as a function of chain length $N$ and
grafting density $\sigma$, assuming good solvent conditions. A
phenomenological scaling theory, describing the variation of the 
linear dimensions of the chains with $\sigma$, is developed and tested by
Molecular Dynamics simulations of a bead-spring model.

Different regimes are identified, depending on the ratio of $D$ to 
the size of a free polymer $N^{3/5}$.  For $D >N^{3/5}$ a crossover 
occurs for $\sigma = \sigma ^* = N^{-6/5}$ from the "mushroom" behavior
($R_{gx}=R_{gy}=R_{gz}=N^{3/5}$) to the behavior of a flat brush
($R_{gz}=\sigma^{1/3}N,\; R_{gx}=R_{gy}=\sigma^{-1/12}N^{1/2}$),
until at $\sigma^{**}=(D/N)^3$ a crossover to a compressed state
of the brush, $(R_{gz}=D,R_{gx}=R_{gy}=(N^3D/4\sigma)^{1/8}<D)$, occurs. 
Here coordinates are chosen so that the $y$-axis is parallel to the tube
axis, and the $z$-direction normal to the wall of the pore at the
grafting site. 

For $D<N^{3/5}$, the coil structure in the dilute
regime is a ''cigar'' of length $R_{gy}=ND^{-2/3}$ along the tube
axis. At $\sigma ^*=(ND^{1/3})^{-1}$ the structure crosses over to
``compressed cigars'', of size $R_{gy}=(\sigma D)^{-1}$. While for
ultrathin cylinders $(D<N^{1/4})$ this regime extends up to the
regime where the pore is filled densely $(\sigma =D/N)$, for
$N^{1/4}<D<N^{1/2}$ a further crossover occurs at $\sigma
^{***}=D^{-9/7}N^{-3/7}$ to a semidilute regime where
$R_{gy}=(N^3D/4\sigma )^{1/8}$ still exceeds $D$. For moderately
wide tubes $(N^{1/2} <D<N^{3/5}$) a further crossover occurs at
$\sigma ^{****} = N^3D^{-7}$, where all chain linear dimensions
are equal, to the regime of compressed brush.

These predictions are compared to the computer simulations. From
the latter, also extensive results on monomer density and free
chain end distributions are obtained, and a discussion of
pertinent theories is given. In particular, it is shown that for
large $D$ the brush height is an increasing function of $D^{-1}$.
\end{abstract}

\section{Introduction}
Polymer brushes are layers of flexible linear polymers on
substrate surfaces produced by endgrafting one chain end via a
special chemical endgroup of the macromolecule to the surface
\cite{1,2,3,4,5}. Such layers have numerous applications in the
fields of colloid stabilization, lubrication, wetting, adhesion,
etc. and hence they have found abiding interest. Polymer brushes
also pose challenging theoretical problems, since the confinement
due to the grafting at the substrates causes the emergence of many
distinct characteristic lengths needed even to describe the
structure of a single chain in a brush \cite{6,7,8}, and although
the density profile of the monomers in the direction normal to the
grafting surface has been studied since a long time ago
\cite{9,10,11,12,13,14,15,16,17}, the degree to which this profile
is understood quantitatively still is a subject of current
discussion \cite{18}.

Note that the overwhelming part of the very rich literature on
polymer brushes (see \cite{1,2,3,4,5} for reviews) deals with
brushes on planar flat surfaces only. Occasionally the problem of
brushes on the outside of spherical particles \{e.g.
\cite{19,20,21}\} and on the inside of spherical cavities
\cite{20,22,23} was considered (as well as weakly curved flexible
membranes, e.g. \cite{24,25}).

However, relatively little work has been done on grafted flexible
polymers inside cylindrical pores (apart from a few comments in
\cite{20,23} and a study of a single grafted chain in a tube
\cite{26}, a problem that is closely related to the behavior of
single non-grafted chains confined in tubes \cite{27,28}). We note
that small mushrooms grafted to cylindrical pores were considered
as a means to control electroosmotic flow \cite{29}.

In the present paper we intend to contribute to fill this gap, by
presenting a study of polymer brushes grafted on the inside of
cylindrical pores of radius R, varying also the chain length N and
grafting density $\sigma$ over a wide range. Such systems should
be possibly also relevant in a biophysical context (cylindrical
pores in biological membranes, blood vessels or other transport
pipes or tube-like objects in biosystems, in which proteins or
other biopolymers can be adsorbed), but this is outside of
consideration here. Rather this paper is devoted to an
understanding of the characteristic lengths that describe the
configurations of the endgrafted chains in such a cylindrical
confinement. In Sec. II, we formulate a (somewhat qualitative)
scaling description, pointing out also the relation to the problem
of non-adsorbed chains in cylindrical pores at various volume
fractions of the monomers \cite{27,30,31,32,33,34}. In Sec. III,
we describe the model used for our simulations to test this
theory, while Sec. IV presents our numerical results for the
linear dimensions of the chains. Sec. V describes results for the
monomer density distribution across the channel, as well as
corresponding results for the distribution function for the free
chain end. Sec. VI then summarizes our conclusions.

\section{Theoretical background: Scaling considerations}

A very brief discussion of polymer brushes in cylindrical geometry
was given in Appendix C of \cite{23}, however, many of the
characteristic lengths relevant for our study were not given, and
also a detailed derivation of the diagram describing the different
regimes in the plane of variables R and $\sigma$ was not
presented; we give such a justification here, in order to make
clear which underlying assumptions are made, and in order to be
able to derive the new results on various characteristic lengths
that are of interest here.

\subsection{Mushrooms and Polymer Brushes on Flat Substrates}
\begin{figure}
\centerline{\includegraphics[width=4.4in, angle=0]{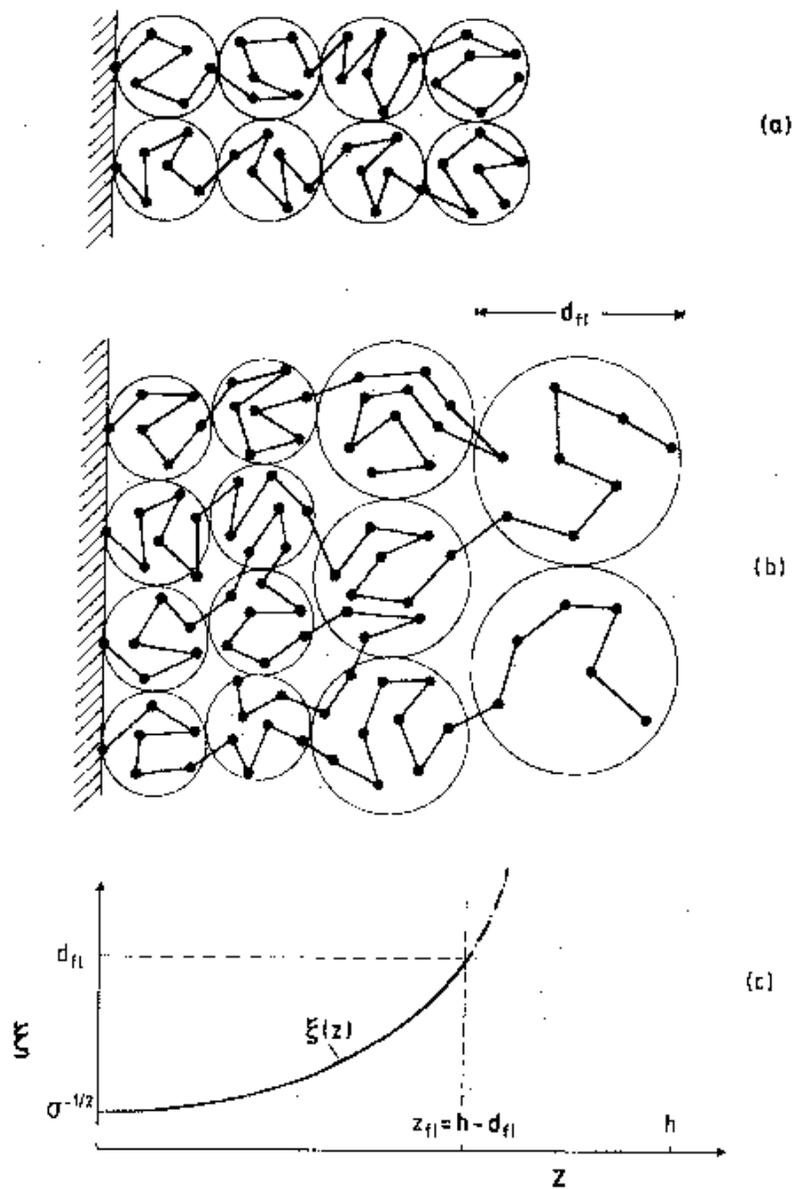}
}
\vspace*{8pt}

\caption{Blob picture of a polymer brush grafted
on a flat substrate in the good solvent regime. (a) Alexander-de
Gennes model: the chains are treated as linear ``cigars'' of blobs
with uniform diameter $\xi = \sigma^{-1/2}$, $\sigma$ being the
grafting density \cite{9,10}. (b) Non-uniform blob picture
\cite{6}: consistent with a decrease of the monomer density $\phi
(z)$ in the brush with increasing distance z from the grafting
substrate surface, the blob diameter (which can also be
interpreted as the screening length $\xi$ of excluded volume
interactions, see (c)) increases with the distance z.
\label{fig1}}
\end{figure}

To set the scene, we recall the ``blob model'' of Alexander
\cite{9} and de Gennes \cite{10} and its extension to deal with a
nonuniform monomer concentration within a brush \cite{6}, see
Fig.~\ref{fig1}. For simplicity, all lengths will be measured in
units of the effective segment length a, so the end-to-end
distance of an ideal (Gaussian) chain containing N segments simply
is $\sqrt{N}$, and both grafting density $\sigma$ and volume
fraction $\phi$ are treated as dimensionless quantities. Only the
good solvent regime is treated, and - as usual in the spirit of
scaling approaches \cite{35} - prefactors of order unity are
suppressed throughout. Also the exponent $\nu$ describing the
linear dimensions of a self-avoiding walk in $d=3$ dimensions is
simply approximated at the Flory value \cite{36} $\nu =3/5$,
rather taking the more accurate value $\nu \approx 0.588$
\cite{37,38}. Thus an isolated polymer chain endgrafted at a
planar wall, a ``polymer mushroom'', has the same gyration radii
components as in the bulk,
\begin{equation}\label{eq1}
R_{gx}=R_{gy}=R_{gz}=N^{3/5}.
\end{equation}
Here, and in the following, we orient the z-axis normal to the
grafting surface while the x and y-axes lie in the grafting plane
(in the case of a a planar brush).

Considering then grafting densities $\sigma$ large enough such
that the individual mushrooms would significantly overlap, we
recognize that (for $\sigma \ll 1$) we obtain a semidilute polymer
brush, in which the screening length $\xi$ is of the order of
$\sigma ^{-1/2}$, which then is taken as the blob-radius in
Fig.~\ref{fig1}. It is implied, that inside a blob excluded volume
statistics persists, only n monomers from one chain are contained
inside of such a blob, and no monomers of any other chains. Since
$\sigma^{-1/2}=n^{3/5}$, in analogy to Eq.~(\ref{eq1}), we
conclude that a blob contains $n=\sigma^{-5/6}$ monomers, and
every chain contains $N_{blob}=N/n=\sigma^{5/6}N$ blobs. According
to the Alexander \cite{9}-de Gennes \cite{10} picture, the chains
in the brush form ``cigars'', i.e. they form linear chains of
blobs stretched out along the z-axis (fig.~\ref{fig1}, upper
part), and hence the brush height $h$ is of the same order as
$R_{gz}$, namely
\begin{equation}\label{eq2}
h=R_{gz}=\xi N_{blob}=\sigma ^{1/3}N.
\end{equation}
Taken literally, this ``cigar'' configuration would imply that
$R_{gx}=R_{gy}=\xi=\sigma^{-1/2}$, which in reality is not the
case: rather the chains in x and y-direction have a structure
resulting from a random walk of blobs, i.e.
\begin{equation}\label{eq3}
R_{gx}=R_{gy}=\xi N_{blob}^{1/2}=\sigma^{-1/12}N^{1/2}.
\end{equation}
While the Alexander \cite{9}-de Gennes \cite{10} picture also
implies that the monomer density profile $\phi(z)$ is essentially
uniform for $z<h$, and the free chain ends are all in the
outermost blob, i.e. in the region $h-\xi<z<h$, the structure in
reality is much more complicated
\cite{6,7,8,12,13,14,15,16,17,18}. Rather than staying constant,
$\rho(z)$ smoothly decreases with increasing $h$ (and in the
strong stretching limit approximately is described by a parabolic
variation, $\phi(z)=\phi(0)[1-z^2/h^2]$). Due to this decrease of
the monomer fraction $\phi(z)$ with the distance z from the
grafting surface, also the screening length $\xi$ is not a
constant, but increases according to $\xi(z)=[\phi(z)]^{-3/4}$
\cite{35}. From a self consistency consideration one can then
estimate the diameter $d_{fl}$ of the final blob as \cite{6}
$d_{fl}=N^{2/5}$. All these concepts have been verified by
corresponding computer simulations \cite{6,7,15,16,17,18}.

\subsection{Cylindrical Tubes}
We now consider a cylindrical tube of diameter D and length L. We
orient the tube axis in y-direction and choose in this direction
periodic boundary conditions.

We first note that there is a simple geometric relation between
the grafting density $\sigma$ of anchor points at the cylinder
surface and the average volume fraction $\phi$ of monomers
contained in the cylinder. Since the area of the cylinder surface
is $A=\pi DL$, the total number of monomers ${\mathcal{N}}$
contained in the cylinder is
\begin{equation}\label{eq4}
{\mathcal{N}} =N\sigma A =N\sigma \pi DL.
\end{equation}
The cylinder volume being $V=\pi LD^2/4$, we find
\begin{equation}\label{eq5}
\phi = {\mathcal{N}}/V=4N\sigma /D
\end{equation}
The condition that the maximum volume fraction that can be reached
in a cylinder corresponds to a dense melt, i.e. $\phi =1$, implies
the following constraint for physically meaningful grafting
densities $\sigma$
\begin{equation}\label{eq6}
\sigma <D/(4N).
\end{equation}
We first discuss the regime of extremely low grafting densities.
For wide tubes, $D >N^{3/5}$, we still have mushrooms, i.e. Eq.~(\ref{eq1})
still holds. A more interesting regime occurs for narrow tubes, 
\begin{equation}\label{eq7}
1<D<N^{3/5},
\end{equation}
where the chains form ``cigars'' stretched along the y-axis
(Fig.~\ref{fig2}). The situation for $\sigma \rightarrow 0$
closely resembles the case of non-grafted chains confined to very
narrow tubes under good solvent conditions \cite{27,28}. 
One has $R_{gx}=R_{gz}=D$, while the longitudinal $y$-component $R_{gy}$ is much
larger. Assuming that the ``cigar'' configuration can be treated
as an array of $N_{blob}=N/n$ blobs of diameter $D$, with $n$ monomers
per blob, $D=n^{3/5}$, we readily obtain
\begin{equation}\label{eq8}
R_{gy}=DN_{blob}=DN/n=ND^{-2/3}
\end{equation}
As it should be, for $D=N^{3/5}$ this yields a smooth crossover to
$R_{gy}=N^{3/5}$.

Next we estimate how small the grafting density $\sigma$ has to be
in order that Eqs.~(\ref{eq1}) or (\ref{eq8}) hold, respectively.
The corresponding boundary in the $D-\sigma$ plane is obtained by
the condition that the grafting density is such that the
mushrooms or cigars just touch each other. 

In the case of
mushrooms, $D>N^{3/5}$, this boundary is reached for an overlap grafting
density $\sigma ^*$ given by
\begin{equation}\label{eq9}
\sigma ^* =N^{-6/5}, \quad D >N^{3/5}
\end{equation}
since each mushroom covers an area of order $N^{6/5}$ when one
projects the monomer coordinates of the chain onto the grafting
surface. 

In the case of cigars, elongated along the $y$-axis, $D<N^{3/5}$, we
simply have to equate the density $\phi$ in the tube to the
density inside a cigar, using Eq.~(\ref{eq8})
\begin{equation}\label{eq10}
\phi ^*=N/(D^2R_{gy})=D^{-4/3}.
\end{equation}
This overlap density $\phi^*$ is easily converted into an overlap
grafting density $\sigma ^*$ using Eq.~(\ref{eq5}),
\begin{equation}\label{eq11}
\sigma ^*=D\phi ^*/(4N)=1/(4ND^{1/3}),\; 1<D<N^{3/5}.
\end{equation}
Also in this case we note that for $D=N^{3/5}$ both expressions
Eqs.~(\ref{eq9}), (\ref{eq11}) for the overlap grafting density
are of the same order, as it should be.

We now consider the behavior for $\sigma > \sigma ^*$. In the
regime where $D>N^{3/5}$ a crossover from mushroom to brush
takes place, similar to the case of planar grafting surface. To
leading order, Eq.~(\ref{eq2}) still holds. However, a strong
compression of the brush sets in when the brush height $h$ (and thus
also $R_{gz}$) become comparable to the tube diameter. This
yields a crossover grafting density $\sigma ^{**}$ separating a weakly
compressed brush (described still by Eq.~(\ref{eq2})) from a
strongly compressed brush, with $R_{gz}=h=D$, for $\sigma >\sigma
^{**}$,
\begin{equation}\label{eq12}
\sigma ^{**}=(D/N)^3
\end{equation}
In the regime of the compressed brush, the density $\phi$ of
monomers is approximately uniform throughout the tube. Then the
lateral linear dimensions are, similar as for chains in a
semidilute solution in the bulk ($\xi= \phi^{-3/4})$, given as
\begin{equation}\label{eq13}
R_{gx}=R_{gy}=\xi N^{1/2}_{blob}=\phi^{-1/8}N^{1/2}=(4
\sigma)^{-1/8}D^{1/8}N^{3/8}, \sigma >\sigma ^{**}
\end{equation}
At the boundary, $\sigma = \sigma ^{**}$, Eq.~(\ref{eq13}) yields
linear dimensions of order
\begin{equation}\label{eq14}
R_{gx}=R_{gy}=D^{-1/4}N^{3/4}\;,\sigma = \sigma ^{**},
\end{equation}
and hence there occurs then again a smooth crossover to
Eq.~(\ref{eq3}).

Finally, we address the behavior for $\sigma > \sigma ^*$ in the
narrow tube regime $D<N^{3/5}$. While for $\sigma = \sigma ^*$ we have found
``cigars'' (Fig.~\ref{fig2}) as described by Eq.~(\ref{eq8}) that
just touch at their ends, we now expect ``compressed cigars'' in
which the excluded volume interactions are partially screened. The
polymer chains overlap themselves and near the cigar ends,
neighboring cigars overlap each other. 
The screening length (blob radius) $\xi$ then is smaller than
the tube diameter, unlike the situation depicted in
Fig.~\ref{fig2}, namely
\begin{equation}\label{eq15}
\xi=\phi^{-3/4}=(4N\sigma / D)^{-3/4},
\end{equation}
where in the last step Eq.~(\ref{eq5}) was used.

For the range of grafting densities $\sigma ^* < \sigma <\sigma
^{***}$ $(\sigma ^{***}$ will be estimated below) we have
compressed cigars with linear dimensions $R_{gx}=R_{gz}=D$ and
linear dimensions $R_{gy}>D$ along the y-axis. Assuming that the
volume $D^2R_{gy}$ is of the same order as the total volume taken
by all the blobs of a chain, $\xi ^3N_{blob}=N\xi^{4/3}$, one finds
\begin{equation}\label{eq16}
R_{gy}=D^{-2}N\xi ^{4/3}=D^{-2}N\phi ^{-1}=1/(4\sigma D)
\end{equation}
The result $R_{gy}=D^{-2}N\phi^{-1}$ is identical to the
corresponding result for non-grafted chains confined to narrow
tubes \cite{30,34}, as expected. In this quasi-one-dimensional
situation the ``cigars'' are laterally compressed, each chain
overlaps itself to some extent (excluded volume is respected only
up to the length $\xi$, Eq.~(\ref{eq15})), but neighboring chains
are still essentially segregated from each other and overlap
only near the ends of the ``cigar''.

Now the boundary $\sigma^{***}$ of this regime can simply be
estimated from the condition that $R_{gy}$, Eq.(\ref{eq16}), equals the result in
the compressed brush regime, Eq.~(\ref{eq13}), which yields
\begin{equation}\label{eq17}
\sigma ^{***} = \frac 1 4 N^{-3/7} D^{-9/7}
\end{equation}
Note that for $D=N^{3/5}$ this boundary starts at $\sigma
^{***}=(1/4)N^{-6/5}$, i.e. it splits off from the boundary
$\sigma ^*$ in the $(D,\sigma)$ plane (Fig.~\ref{fig3}). The other
end of this boundary is reached for the condition $\phi =1$, i.e.
$4\sigma ^{***}=D/N$ (Eq.~(\ref{eq5})) and hence $D=N^{1/4}$. This
means, for ultra-narrow tubes (defined by the condition
$D<N^{1/4}$) Eq.~(\ref{eq16}) holds from $\sigma = \sigma ^*$ up
to $\sigma = D/4N$, so the boundary $\sigma^{***}$ cannot be
reached.

For tubes that are not ultra-narrow $(D > N^{1/4})$ one encounters
at $\sigma ^{***}$ a crossover to a regime of semi-diluted
cigars where now two of the three linear dimensions
$R_{gx}=R_{gz}=D$, while the third linear dimension, $R_{gy}$, is
given by Eq.~(\ref{eq13}). However, the cigars now start to
overlap more and more, until at a grafting density $\sigma
^{****}$ the ``natural size'' $R_{gy}$ of a chain in a semidilute
solution of density $\phi$ reaches $D$,
\begin{equation}\label{eq18}
R_{gy}=\xi N_{blob}^{1/2}=\phi^{-1/8}N^{1/2}\;,\quad
\sigma^{***}<\sigma<\sigma^{****}.
\end{equation}
While in the regime described by Eq.~(\ref{eq18}) $R_{gy}$ still
exceeds D, and therefore the linear dimension of a chain in
$y$-direction exceeds the other linear dimensions, we denote this
regime as ``overlapping cigars''. The grafting density $\sigma
^{****}$ is found from the condition that $R_{gy}=D$, and hence
\begin{equation}\label{eq19}
D=\xi N_{blob}^{1/2}=(4N\sigma
/D)^{-1/8}N^{1/2}\;,\;\sigma^{****}= \frac 1 4 N^3D^{-7}
\end{equation}
For $\sigma ^{****}< \sigma < D/4N$ we enter the same region of
compressed brush that we have already encountered for $D>N^{3/5}$
for $\sigma ^{**} < \sigma <D/4N$. The line $\sigma ^{****}$
starts in the $(D,\sigma)$ plane also for $D=N^{3/5}$ at the point
$\sigma = \frac 1 4 N^{-6/5}$, i.e. at $\sigma^*$, and ends for
$\phi = 1(\sigma=D/4N)$ at the diameter $D=N^{1/2}$
(Fig.~\ref{fig3}).

The behavior of the brush in a tube becomes somewhat more
transparent if we discuss it in terms of the scaled variables
$\tilde{D} \equiv D/N^{3/5}, \tilde{\sigma} \equiv \sigma N
^{6/5}$, since then (in the dilute and semidilute regime) the
dependence on the chain length is absorbed completely.
Fig.~\ref{fig4} shows qualitatively the dependence of the chain
linear dimensions: For $\tilde{D}>1$ the scaled radius
$\tilde{R}_{gz} \equiv R_g/N^{3/5}$ crosses over from unity at
$\tilde{\sigma}=1$ to the power law $\tilde{\sigma}^{1/3}$ and at
$\tilde{\sigma}= \tilde{\sigma}^{**}= \tilde{D}^3$ to $\tilde{D}$,
while the other components of the radius,
$\tilde{R}_{gx}=R_{gx}/N^{3/5},\; \tilde{R}_{gy}=R_{gy}/N^{3/5}$
cross over from unity at $\tilde{\sigma}=1$ to the power law
$\tilde{\sigma}^{-1/2}$ and at $\tilde{\sigma}^{**}$ (where
$\tilde{R}_{gy}=\tilde{D}^{-1/4}$) a further crossover to
$(\tilde{D}/\tilde{\sigma})^{1/8}$ occurs.

Similarly, for $\tilde{D}<1$ the radius $\tilde{R}_{gy}$ is
$\tilde{D}^{-2/3}$ for small $\tilde{\sigma}$ and crosses over at
$\tilde{\sigma}^* =\tilde{D}^{-1/3}$ to a power law
$(\tilde{\sigma}\tilde{D})^{-1}$. For $\tilde{D}<N^{-4/15}$ this
is the only crossover that occurs, while for larger $\tilde{D}$
additional crossovers are found at $\tilde{\sigma}^{***}$ (and
$\tilde{\sigma}^{****}$, if $\tilde{D}$ exceeds $N^{-1/10}$) with
$\tilde{\sigma}^{***}=\tilde{D}^{-2},\;\tilde{\sigma}^{****}=\tilde{D}^{-7}$.

For $\tilde{\sigma}>\tilde{\sigma}^{***}$ we have
$\tilde{R}_{gy}=\tilde{D}$, but only at $\tilde{\sigma} =
\tilde{\sigma}^{****}$ the radius of a chain in a bulk
semidilute solution of density $\tilde{\phi}$ corresponding to
$\tilde{\sigma}$ would be of order $\tilde{D}$ and one goes over
to compressed brush behavior where $\tilde{R}_{gy}$
decreases to linear dimensions smaller than $\tilde{D}$, namely
$\tilde{R}_{gy}=(\tilde{D}/\tilde{\sigma})^{1/8}$, as above. 

Note that in the case $\tilde{D}<1$ the behavior of the brush in the
tube in many aspects is similar to the behavior of non-grafted
chains in a tube, as discussed by \cite{30,34}. One can see this
converting $\sigma$ to $\phi$, using Eq.~(\ref{eq5}); however, for
the latter problem the crossover at $\tilde{\sigma}^{****}$ has a
different meaning: then all three linear dimensions
$\tilde{R}_{gx},\; \tilde{R}_{gy},\;\tilde{R}_{gz}$ are smaller
than $\tilde{D}$, given by $\tilde{R}_{g\alpha} =
(\tilde{D}/\tilde{\sigma})^{1/8}$, for $\alpha = x,y,z$ while for
$\tilde{\sigma} <\tilde{\sigma}^{****}$ { we have
$\tilde{R}_{gx}\equiv\tilde{R}_{gz} = \tilde{D}$ for the
non-grafted confined chains. For the grafted confined chains, we
also have $\tilde{R}_{gx}= \tilde{R}_{gz}=\tilde{D}$ in this
regime, but $\tilde{R}_{gx}$ and $\tilde{R}_{gz}$ may differ from
each other by factors of order unity (which are suppressed here in
our scaling description), while for the non-grafted chains these
components are identical by the symmetry of the problem. For
$\tilde{\sigma}>\tilde{\sigma}^{****}$, however, there is a
difference on a scaling level, since the grafting of the chain
ends at the walls creates an elastic energy in the brushes,
causing chain stretching in the normal direction $(z)$, and
$\tilde{R}_{gz}$ stays at its (maximal) value which is of order
$\tilde{D}$. Finally, we note that in the scaling diagram of
Manghi et al. \cite{23} the line corresponding to
$\tilde{\sigma}^{***}$ was not included. From Fig.~\ref{fig4},
however, it is evident that an intermediate regime between the
isolated cigars ($\tilde{R}_{gy}=\tilde{D}^{-2/3}$) and the
semidilute regime
$(\tilde{R}_{gy}=(\tilde{D}/\tilde{\sigma})^{1/8}$) is mandatory
to provide a smooth crossover of the linear dimension
$\tilde{R}_{gy}$ and $\tilde{\sigma}^*$. Note also, that on the
crossover line $\tilde{\sigma}^{****}$ where
$\tilde{R}_{gy}=\tilde{D}$ and hence all linear dimensions are of
order $\tilde{D}$, the behavior of $\tilde{R}_{gx}$ changes: we
have $\tilde{R}_{gx}=\tilde{D}$ for
$\tilde{\sigma}<\tilde{\sigma}^{****}$ but
$\tilde{R}_{gx}=\tilde{R}_{gy}=(\tilde{D}/\tilde{\sigma})$ for
$\tilde{\sigma} >\tilde{\sigma}^{****}$.

\section{Model and some Remarks on Simulation Methods}
Being interested in phenomena of mesoscopic length scales of order
$10 nm$ (or more) where chemical detail only enters into
prefactors of various quantities, and a scaling description as
developed in Sec. II makes sense, the use of a coarse-grained
model where atomistic details are not included is mandatory for
the sake of computational efficiency of our computer
simulation\cite{39,40,41,42,43}. There is compelling evidence that
the generic properties of flexible polymers in solution and melt
can be described by simple bead-spring models where each bead may
represent  a few chemical monomers along the chain
backbone\cite{39,47}, and actually such models have been used for
previous simulations of polymer brushes\cite{3,4,15,18,48,49}
successfully.

In this bead-spring model the interactions between the beads
modelling the monomer units are described by a Lennard-Jones (LJ)
potential that is truncated at $r_c$ and shifted there to zero,
\begin{eqnarray}\label{eq20}
U_{LJ}(r)=4\varepsilon \left [\left (\frac{\sigma_{mm}}{r}\right
)^{12}-\left (\frac{\sigma_{mm}}{r} \right )^{6}\right
]+\varepsilon \frac{127}{4096}, \; r<r_c \label{LJ}
\end{eqnarray}
where $r_c=2^{7/6}\sigma_{mm}$. Here $\varepsilon$ characterizes
the strength and $\sigma_{mm}$ the range of the LJ-potential. Note
that the minimum of $U_{LJ}$ occurs at
$r_{min}=2^{1/6}\sigma_{mm}$, without the additive constant
$\varepsilon \frac{127}{4096}$ the depth of this minimum would be
$U_{LJ}(r=r_{min})=-\varepsilon$. Due to the shift of $U_{LJ}$ at
$r_c=2r_{min}$, the potential in Eq.~(\ref{LJ}) is continuous at
all $r > 0$.

In addition, monomers bonded to each other as nearest neighbors
along a polymer chain interact with a finitely extensible
nonlinear elastic (FENE) potential \cite{44}
\begin{eqnarray}\label{eq21}
U_{FENE}(r)=-15\varepsilon
\left(\frac{R_0}{\sigma_{mm}}\right)^2\log\left(1-\frac{r^2}{R_0^2}\right),
\; R_0=1.5\sigma_{mm}. \label{FENE}
\end{eqnarray}
The choice of these parameters ensures that the minimum of the
total potential between two bonded monomers along the chain occurs
for \cite{45,46} $r\approx 0.96\sigma_{mm}$, distinct from
$r_{min}\approx 1.12\sigma_{mm}$. This misfit between the two
distances ensures that there is no tendency of monomers to form a
simple crystal structure even if the density is very high and/or
the temperature is rather low. However, in the present paper we
shall consider only temperatures in the good solvent regime of our
model. This means, the temperature $T$ must exceed the
$\Theta$-temperature, which for the present model has been
established as \cite{47} $k_BT/\epsilon \approx 3.3$. Thus, we
choose $k_B T /\epsilon = 4$ for our simulation. To simplify the
notation we choose $k_B \equiv 1$ and take $\varepsilon \equiv 1$
and $\sigma_{mm}=1$ as our units of energy and length,
respectively.

The grafting wall is represented by atoms forming a triangular
lattice, wrapped around a torus by a periodic boundary condition.
This choice of wall is inspired by carbon nanotubes. However, the
potential with which effective monomers and wall atoms interact is
chosen to be the same  as the monomer-monomer potential,
Eq.~(\ref{LJ}). Also, the wall atoms are fixed in the rigid
positions of an ideal crystal lattice.  This guarantees that no
monomer  can cross the wall of the tube and get out of it. Of
course, the coarse-grained nature of our polymer model would make
a chemically realistic description of a cylindric pore
meaningless. The polymers are  grafted hexagonally on a cylinder
with radius $R - 1.0$, thus defining a grafting density $\sigma$
while the tube atoms are placed a little further - on a cylinder
with radius $R$. It might be more realistic, of course, to choose
randomly distributed grafting sites, however, the experience with
flat brushes\cite{17} shows that it makes little difference
whether the grafting sites are distributed regularly or at random.
The size of the simulation box along its axis $y$ is $L_y$ and the
number of static particles, forming the tube $N_{tube}$, are
different for simulation of tubes with different radii $R$: for $R
= 15$,  $L_y = 81.62$,  $N_{tube} = 1024$; for $R = 21$,  $L_y =
76.18$, $N_{tube} = 2400$; and for $R = 31$: $L_y = 67.47$,
$N_{tube} = 1240$.

Simulations are performed by a standard leapfrog algorithm at
constant temperature, maintained by a Nos\`e-Hoover thermostat \cite{48},
combined with a simple ``hot spot control'' algorithm which
rescales the velocity of every monomer if the current ``particle
temperature'' differs from the average value too much. Choosing
the mass of an effective monomer as $m\equiv 1$, the MD time unit
is $t_0=(\sigma_{mm}^2 m /48\varepsilon)^{1/2}$, and the time step
of integration $\delta t = 0.0005t_0$.

Particular care has to be devoted to the construction of the
initial configuration of the grafted polymers and their
equilibration. The construction of the starting configuration for
given $(N, \sigma, R)$ begins with the growth
 of the chains in a tube with radius $R_{init} = 2R$ and $N = N_{max}$, where
$N_{max}$ is the number of monomers in the longest chain,
investigated for these $\sigma$ and $R$. After placing the first
monomer at the grafting site, each consecutive monomer is placed
at a distance $0.75 R_0$ from the preceding one along a direction
${\bf r} = 2.0* {\bf e_{0z}} + {\bf e_{rand}}$ where ${\bf
e_{0z}}$ is a unit vector, pointing radially from the grafting
site to the tube axis, and ${\bf e_{rand}}$ is a unit vector
oriented at random. After an initial equilibration of this system
for about $10^5$ MD steps, we rescale the system radius to the
final system with desired radius $R$ and again let the system
equilibrate for $10^5$ steps. From the final configuration of this
system, by cutting systematically the polymer chains to different
values of $N$, we obtain the ``starting configurations'', which we
additionally equilibrate for about $3.10^5$ steps and (depending
on $N$ and $\sigma$) then collect statistics during the following
$2.10^5 - 4.10^5$ integration steps. Typical examples of
equilibrated brush configurations in this cylindrical geometry are
shown in Fig.~\ref{fig5}. For all three tube diameters $D=30, 42,$
and $62$ as well as for chain lengths $8\le N \le 64$ we then
sample brush profiles and components of gyration radius, $R_g$,
and end-to-end distance vectors of the chains, $R_e$, for seven
different grafting densities $\sigma$. In our simulations we have
refrained, however, from creating elongated cigar-like
configuration, cf.~Fig.~\ref{fig2}b, since the distribution
$\rho(y)$ of $R_e$ is expected to be bimodal (that is, the
center-of-mass of the chains is either to the left, or to the
right of the grafting site, as shown in Fig.~\ref{fig2}b). The
equilibrium $\rho(y)$ is expected to produce a pronounced minimum
above the grafting site at $y=0$ and, therefore, a well
equilibrated distribution $\rho(y)$ would imply that the chain end
passes the coordinate $y=0$ many times which can hardly be
possible within the typical time interval of a MD simulation.

\section{Linear dimensions of chains in brushes in cylindrical
pores} 

In order to test our model and check that for the rather
short chain lengths used in the present work (typically $8 \leq N
\leq 64$) results are obtained that one can compare to the
theoretical predictions, which really address the asymptotic
regime $N \rightarrow \infty$ only, we first present data for
brushes on flat substrates in Figs.~\ref{fig6}, \ref{fig7}. One
recognizes that for small $\sigma$, such as $\sigma \approx
0.0115$, the linear dimensions in the $z$-direction, perpendicular
to the substrate, and in the $y$-direction, are rather similar for
the range of chain lengths that is studied, but the effective
exponents $\nu_{eff}$ in the relations $\langle R_{gy}^2\rangle
\propto N^{2\nu_{eff}},\quad \langle R_{gz}^2\rangle \propto N^{2
\nu_{eff}}$ somewhat differ from the theoretical value in
the mushroom regime, $\nu = 3/5$ (Eq.~(\ref{eq1})). For the
grafting density $\sigma \approx 0.1039$, however, we recognize
that the scaling behavior of the chain linear dimensions 
strongly depends on direction,
$\langle R_{gz}^2 \rangle \propto N^{1.86}$ while
$\langle R_{gy}^2\rangle \propto N^{1.16}$, 
and approaches the theoretical behavior that
is expected for much larger N, $\langle R_{gx}^2\rangle
\propto N$ while $\langle R_{gz}^2\rangle \propto N^2$
(Fig.~\ref{fig6}c). Since comparatively short chains are used, it would be
premature to expect that one reproduces the asymptotic exponents
of the scaling theory precisely. Also the dependence of both
$\langle R_{gy}^2\rangle$ and $\langle R_{gz}^2\rangle$ on $\sigma$ is roughly
compatible with the predicted power laws (Eqs.~(\ref{eq2}),
(\ref{eq3})). Of course, for polymer brushes on flat substrates it
would be easier to equilibrate
significantly longer chains: e.g., \cite{48} used $50 \leq N \leq
200$ for a closely related model, and \cite{18} used $16 \leq N
\leq 512$. Even then difficulties to verify scaling were
encountered \cite{18}.

However, for the brushes in the cylindrical pores good
equilibration is significantly more difficult; therefore we
restrict the present work to relatively short chains.

Thus, it cannot be a surprise that in the scaling plots
(Figs.~\ref{fig7}a,b) a perfect collapse on master curves is not
obtained, and rather some systematic deviations due to corrections
to scaling are clearly apparent. Such corrections to scaling are
presumably also responsible for the fact that the effective
exponents $\nu_{eff}$ in the mushroom regime of
Figs.~\ref{fig6}b,c deviate from their theoretical value. Note
also that in the present work a somewhat more restricted range of
grafting densities is studied than in \cite{18}, where much lower
grafting densities were included to ensure to reach the mushroom
regime fully. Our data in Fig.~\ref{fig6} even for the smallest
values of $\sigma$ are already affected a little bit by the
crossover towards the brush regime, since this crossover occurs
rather gradually.

In \cite{18} also the problem was discussed that for temperatures
not too far from the Theta temperature the excluded volume
interaction on the scale of a blob diameter may be so weak that
then in the regime of high grafting densities inside of a blob
gaussian chain statistics still holds. As a consequence, 
then a further crossover 
that occurs for large $\tilde{\sigma}$ under good solvent conditions
has been identified. However, this crossover gives rise to rather
small and subtle effects, and hence can be safely ignored in the
present context.

We now turn to our data for brushes in cylindrical pores.
Figs.~\ref{fig8}-\ref{fig10} present the linear dimensions
$\langle R^2_{gz}\rangle $ and $\langle R_{gy}^2\rangle $ for
three choices of the pores diameter, $D=30, \;42,$ and 62. For the
narrow pore ($D=30$), one sees a weak increase of $\langle
R_{gz}^2\rangle $ with $\sigma$ for small $N$, while for larger $N$ no
longer any increase occurs. This behavior is in marked contrast
with the behavior of the brush on the flat substrate. In contrast,
$\langle R^2_{gy} \rangle$ decreases with $\sigma$ monotonously,
while the theoretically predicted double crossover
(Fig.~\ref{fig4}) is not seen. For the wider tube, such as $D = 62$,
one still sees a more pronounced increase of $\langle R^2_{gz}
\rangle $ with $\sigma$, but if one would fit a power law to this
increase, the effective exponent would be distinctly smaller than
the theoretical value 2/3, that should still be observable in an
intermediate regime of $\sigma$, see Fig.~\ref{fig4}. However,
since the crossovers (from the mushroom regime to the brush and
from the brush to the compressed brush) do not occur at sharply
defined values of $\sigma$, but are rather expected to be smeared
out over a range of values $\sigma$ that may be a full decade or
even more. Therefore instead of the two ``kinks'' in Fig.~\ref{fig4} a
smooth gradual increase with $\sigma$ is noted. Similarly, when we
study $\langle R^2_{gz} \rangle$ as a function of N
(Fig.\ref{fig10}c), one can still see straight lines on the
log-log plot, but the observed effective exponents are somewhere
in between $2\nu=1.2$ and $2$, indicating the onset of compression
due to the confinement of the brush in the tube.
However, such effective exponent which simply reflect slow
crossovers do lack any deep physical significance, of course.

Motivated by the scaling analysis of Sec.~II, we discuss the
extent to which the chain length dependence can be absorbed in
scaling plots of $\tilde{R}_{gy}$ and $\tilde{R}_{gz}$ versus
$\tilde{\sigma}$ (Fig.~\ref{fig11}). Of course, only for
$\tilde{D}>1$ the first crossover from mushrooms to the almost
unperturbated brush is expected to stay unaffected by $D$. Since
varying $N$ at fixed $D$ means that different values of $\tilde{D}$
occur, the positions of the various other crossovers in
Fig.~\ref{fig11} do depend on $N$ via the $N$-dependence of
$\tilde{D}$. So we expect that the curves for $\tilde{R}_{gz}^2$
vs $\tilde{\sigma}$ at large enough $\tilde{\sigma}$ start to
split off from the master curve that describes both the mushroom
and brush regime and saturate at plateaus corresponding to the
different values of $\tilde{D}$. Some evidence for the splaying
out of the curves on the right hand side of the scaling plot for
$\tilde{R}_{gz}^2$ can indeed be seen in Fig.~\ref{fig11}.
However, for the parameters accessible in our study the crossovers
at $\tilde{\sigma}^*=1$ and at $\tilde{\sigma}^{**}=\tilde{D}^3$
are too close together, so the intermediate regime where
$\tilde{R}_{gz}^2 = \tilde{\sigma}^{2/3}$ is never clearly
realized. In fact, in order to clearly separate these crossovers,
one would need to have not only $N \gg 1$ but also $\tilde{D}\gg
1$, and this has not been achieved. For small $N$ corrections to
scaling come into play also and obscure the behavior further. Part
of our data falls into the region $\tilde{D} <1$, where a
different pattern of crossover behavior applies, but again the
different regions are clearly separated only if $\tilde{D}\ll 1$
is achieved. We could not reach this region due to difficulties of
equilibration.

Thus Fig.~\ref{fig11} does not give strong evidence for the
scaling theory of Sec.~II: clearly these scaling considerations
are useful only for the special limit where both $N$ and $D$ are very
large, $\sigma$ being very small, so that $\tilde{\sigma}$ can be
varied over many decades, as well as $\tilde{D}$. This limit is
not easy to reach in our simulations. Nevertheless the scaling
theory gives a useful orientation: note that the data for the
scaled quantities $\tilde{R}_{gy}^2$ and $\tilde{R}^2_{gz}$ in
Fig.~\ref{fig11} show remarkably little variations in spite of the
fact that both $N$ and $\sigma$ are varied over about an order of
magnitude each.

\section{Monomer density and chain distribution across a
cylindrical pore}

We start this section again with a brief glimpse on the behavior
of our model in the limit of a flat grafting surface - 
Fig.~\ref{fig12}. 
As expected, one sees a smooth
decay of the monomer density profile $\phi(z)$ with increasing
distance z from the surface. When N increases, the height h of the
brush (region where $\phi (z)$ is appreciably nonzero) strongly
increases (Fig.~\ref{fig12}a). Since from
Eq.~(\ref{eq2}) we expect that $h \propto N$, it is interesting to
plot $\phi(z)$ against a rescaled coordinate $z/N$
(Fig.~\ref{fig12}b). One can see that for large $N$ the
curves indeed converge smoothly towards a master curve, and this
master curve resembles the simple parabolic shape predicted by the
selfconsistent field (SCF) theory in the strong stretching limit
\cite{12,13,14}
\begin{equation}\label{eq22}
\phi(z)=\phi(0)[1-(z/h)^2]\; \quad N \rightarrow \infty, \; \sigma
\rightarrow 0, \; N\sigma ^{1/3}\; {\textrm{finite}}
\end{equation}
Eq.~(\ref{eq22}) implies a nonzero intercept $\phi(0)$, while the
actual $\phi (z \rightarrow 0) \rightarrow 0$, of course
(Fig.~\ref{fig12}): so the $\phi(0)$ of Eq.~(\ref{eq22}) is
interpreted as an extrapolation of the flat part of the profile;
also $\phi(z)\equiv 0$ for $z >h$, which for finite $N$ is not true.
However, numerical implementations of SCF theory that do not
invoke the strong stretching limit, such as the Scheutjens-Fleer
theory \cite{51}, can describe the smooth decay of the tail of the
profile $\phi(z)$ for $z>h$ \cite{52}, but this is not attempted
here. In order to provide a precise definition of $h$ from the
observed profiles $\phi(z)$, such as shown in Fig.~\ref{fig12}, we
follow previous practice (e.g. \cite{18}) to use the first moment
$\langle z \rangle$,
\begin{equation}\label{eq23}
h = \frac 8 3 \langle z \rangle,\quad \langle z \rangle = \int
\limits _0^\infty z \phi (z) / \int \limits _0^\infty \phi (z)dz.
\end{equation}
It is easily checked that Eqs.~(\ref{eq22}) and (\ref{eq23}) are
compatible with each other, but Eq.~(\ref{eq23}) is generally
applicable. The insert of Fig.~\ref{fig12} gives evidence that the
scaling regime $h \propto N$ has not yet fully been reached by our
rather short chains, but at least there is evidence that the
chains are stretched away from the surface. Similar data have been
generated by many previous simulations on various models
\cite{3,6,15,16,17,18}. The delta-function like peak for the first
monomer (at $z=1.0)$ simply is due to the fact that the wall atoms
forming the flat surface are at a fixed distance $z=1.0$ away from
the first monomer of the chain.

We also draw attention to the weak oscillations in the density
profile (``layering'') that is visible for small $z$. Such effects
arise from the ``packing'' of monomers near the grafting surface,
which resembles a hard wall. Such packing effects are neither
described by the scaling theory of Sec. II nor the SCF theory, of
course.

Also the distribution function $\rho (z)$ of the end monomers is
of interest (Fig.~\ref{fig12}c). One can see that the
free chain ends are distributed all over the brush, unlike the
prediction of the Alexander \cite{9} - de Gennes \cite{10}
picture, according to which the chain ends are all localized at a
distance $z \approx h$ from the grafting surface. According to the
strong stretching limit of SCF theory, however, one would predict
a distribution that is spread over the entire brush but vanishes
at $z = h$ with a divergent slope \cite{12},
\begin{equation}\label{eq24}
\phi_e(z)\propto z[1-(z/h)^2]^{1/2},\; N\rightarrow \infty,\;
\sigma \rightarrow 0, \; \sigma ^{1/3}N \;{\textrm{finite}}
\end{equation}
From Fig.~\ref{fig12}c it is clear that our data do
not reach the strong stretching limit of the SCF theory, however.

For polymers grafted to a flat wall it would be possible to
simulate much longer chains, of course, and, in fact, in
\cite{18} for a closely related model (perfectly flat wall with no
corrugation whatsoever and using an integrated Lennard-Jones-type
repulsive wall potential) chain lengths up to N = 512 were
studied. However, even for the largest chains the strong
stretching limit of the SCF theory was not reached. Since the main
interest of the present paper concerns chains in cylindrical
tubes, and for this case it is probably not feasible
experimentally to graft extremely long chains at the tube walls,
we shall not emphasize the comparison with the strong stretching
limit of the SCF theory in the present paper; just as the scaling
considerations of Sec. 2 it is useful for a first orientation, but
one cannot rely on its quantitative details in the present case.

After this prelude on the density profiles for the well-studied
case of polymer brushes grafted to planar substrates we turn to
our results on the profiles in cylindrical pores,
Figs.~\ref{fig13}-\ref{fig16}. While for short chains (such as
$N=8,12,16$) and low grafting densities these density profiles are
typical of mushroom behavior, for larger $N$ one quickly reaches a
state where the whole cylindrical brush is more or less uniformly
filled with monomers. In this limit the density of end monomers
reaches a maximum near the tube axis, but in comparison with the
brushes on flat substrates (Fig.~\ref{fig12}) the distribution of
the end monomers is more uniform. Thus the approximation of Sevick
\cite{20} assuming a delta-function distribution of chain ends
(like in the Alexander \cite{9}-de Gennes \cite{10} model)
obviously is not accurate, at least for the parameter range
accessible in our simulations. At this point we recall that Sevick
\cite{20} predicts that the brush height $h$ decreases with
increasing inverse cylinder diameter $D$ monotonously, while Milner
and Witten \cite{53} on the basis of their SCF theory suggest that
the brush height $h$ increases with $1/D$ (for small enough $1/D$).
The interpretation for the latter result is that in the
cylindrical pore the blob diameters get smaller with increasing
distance z from the grafting surface when $1/D$ increases, since
the volume available for the blobs decreases, and hence the
tendency of the brush towards stretching is more pronounced in the
outer region of a brush in a cylinder than for a brush at a flat
substrate. Of course, when h and $D/2$ become comparable, the
opposite tendency of compressing the brush due to the confinement
will outweigh this increase of h that may be present for small
$1/D$.

Again we emphasize that the grafted monomer (adjacent to the wall
of the tube) is at a fixed distance $\Delta r=1$ inside of the
atoms forming the walls of the tube, which are at the radial
distance $r=R$ from the tube axis. The sharp spike at $r=R-1$  in
Figs.~\ref{fig13}a, \ref{fig14}a, \ref{fig15}a, }\ref{fig16}a
hence is always a representation of the delta function
corresponding to the monomer adjacent to the rigid wall, at this
fixed distance $\Delta r=1$ inside of it. However, other monomers
of the chain can come closer to the wall, when the chain bends
back to the grafting wall. Therefore one finds a nonzero $\phi(r)$
in between $r=R-1$ and $r=R$, as is clearly seen in
Figs.~\ref{fig13}a and \ref{fig14}a.

When the chain length $N$ is small enough (and/or the tube diameter
$D$ is large enough), both $\phi (r)$ and $\rho(r)$ are essentially
zero for $r \rightarrow 0$. The pore center then is still free of
monomers, and Figs.~\ref{fig13} - \ref{fig16} then contain also
the full information on the distribution of the monomers as a
function of the radial distance {\em from the grafting wall}, $r'$, we simply
can transform from $r$ to $z$ via $r'=D/2 - r$. However, when either
$N$ gets larger or $D$ gets smaller (or both), these distributions
$\phi (r)$ and $\rho(r)$ get nonzero for $r \rightarrow 0$. This
means that then distances larger than $D/2$ become possible
\{cf. Fig.~\ref{fig2}(a)\}. Then each value of $r$ actually
contains two entries for $r'$, namely
\begin{equation}\label{eq25}
r'=D/2-r\;, {\textrm{for}}\; r'<D/2, \; {\textrm{and}}\; r'=D/2+r, \;
{\textrm{for}}\; r'>D/2.
\end{equation}
The choice $r'=D/2+r$ has to be used when the monomer lies in the
upper hemicylinder with respect to the grafting site taken as the 
lowest point of the lower hemicylinder.
Thus the tube axis in the simulation is at $r'=R=D/2$,
since we arbitrarily defined $r=R-1$ to be the position of the
first monomer and put the wall at the radial distance $r=R$ from
the tube axis.

Because of this ambiguity, it is also of interest to analyze
separately the distribution functions $\phi(r')$ and $\rho (r')$.
Fig.~\ref{fig17} gives an example (more details will be given in a
forthcoming publication). Of course, for the cases which $\phi
(r\rightarrow 0)\rightarrow 0$ and $\rho(r\rightarrow 0$) the
curves plotted in Fig.~\ref{fig17}a are identical to those of
Fig.~\ref{fig13}, only now left and right is interchanged and the
origin of the coordinate system shifted. However, this is not so in
the cases where $\phi(r \rightarrow 0)$ and $\rho(r \rightarrow
0)$ are nonzero: we now see that $\phi (r')$ and $\rho(r')$ look
qualitatively very similar to corresponding curves for brushes at
flat walls. In particular, there is no singular behavior at the
center of the tube, at $r'=R$ all curves are perfectly smooth. Also
the distribution of chain ends $\rho (r')$ is perfectly smooth at
this distance $r'=R$ and carries considerable weight for distances
$r' >R$ if $N$ is sufficiently large (and/or $D$ sufficiently
small), as Fig.~\ref{fig18}b shows.

It is also of interest to examine the distribution of the chain
end $\rho (y)$ as a function of the lateral coordinate $y$,
parallel to the tube axis (Fig.~\ref{fig18}). One can see that the
width of this distribution does not depend much on the grafting
density, and is largest for the smallest grafting density. No
evidence for bimodal distributions, that are expected for narrow
tubes and long enough chains due to the cigar-like configurations
is found for the parameter range investigated, however.

The profiles $\phi (r')$ can be used to calculate the first moment
$\langle r' \rangle$, from $\langle r' \rangle$ one obtains the brush
height $h$, making use of Eq.~(\ref{eq23}). 
Fig.~\ref{fig19} shows a plot of the reduced height, $h/N$, 
versus $N$ for $D=42$ and $D=62$ and two different grafting densities.
A first important conclusion which one may draw upon inspection of Fig. 
\ref{fig19} suggests that the brush height {\em grows} with increasing
curvature of the tube, in contrast to the basic assumption in Sevick's
theory \cite{20}. Evidently, when the tube diameter diminishes
the monomer density grows, cf. the inset of Fig. \ref{fig19} where also
the monomer density profile for a flat substrate, that is, zero curvature, 
is displayed. Clearly, as the density grows, the excluded volume interactions "push" 
the brush height toward the tube axis. As shown in Fig. \ref{fig19},  
this effect is seen irrespective of particular grafting density $\sigma$. 
At higher $\sigma$, however, one can readily see that beyond a certain
chain length ($N=32$ for $D=42$, and $N=40$ for $D=62$) the ratio
$h/N$ reaches a plateau, as expected on the ground of Eq. (\ref{eq2})
whereas for the lower grafting density, $\sigma = 0.03$, the onset of
this scaling behavior probably occurs at much larger chain lengths $N$. 

\section{Conclusions}
The configurations of flexible polymers endgrafted at the inner
walls of cylindrical tubes under good solvent conditions were
studied by two complementary approaches, namely by a scaling
theory and by Monte Carlo simulation of a bead-spring model.

The scaling treatment can identify the asymptotic power laws of
the chain linear dimensions $R_{gx},\; R_{gy}$ and $R_{gz}$ and
the brush height $h$ in terms of the parameters of the problem,
namely the chain length $N$, the grafting density $\sigma$, and
the tube diameter $D$. These power laws, however, are valid only
asymptotically in the limit $N \rightarrow \infty$, keeping the
scaled grafting density $\tilde{\sigma}=\sigma N^{6/5}$ and the
scaled tube diameter $\tilde{D}=DN^{-3/5}$ fixed. In this limit,
also the scaled linear dimensions
$\tilde{R}_{g\alpha}=R_{g\alpha}N^{-3/5} \; (\alpha=x, y,z)$ and
$\tilde{h}=hN^{-3/5}$ will not exhibit any explicit dependence on
$N$ any more, but will depend on the scaled variables
$\tilde{\sigma}$ and $\tilde{D}$ only (see Fig.~\ref{fig4}, for
example).

It should be noted, that even in this limit the scaling theory is
grossly oversimplified, since the different regimes that have been
identified (mushrooms, cigars; compressed cigars; overlapping
cigars; compressed brush; normal brush) are not separated by any
sharp boundaries, but rather the different power laws that
characterize these regimes merge gradually by smooth crossovers.
These crossover regions may extend over several decades of the
appropriate control variable (such as $\tilde{\sigma}$). Therefore
the lines in the ``state diagram'' of the polymer brush
(Fig.~\ref{fig3}) should not be mixed up with lines appearing in a
thermodynamic phase diagram: here the lines only mark the center
of the crossover region (fig.~\ref{fig3}), and all the kinks in a
scaling diagram such as Fig.~\ref{fig4} are rounded off. However,
we are not aware of any analytical method that could provide the
functions that describe these smooth crossovers.

Real polymer brushes confined in such a cylindrical pore geometry
will hardly obey these rather stringent conditions under which the
scaling considerations are strictly applicable, and the same
caveat holds for computer simulations. Nevertheless the scaling
approach is useful as a first guidance towards this rather complex
behavior.

\begin{figure}
\centerline{\includegraphics[width=2.2in, angle=0]{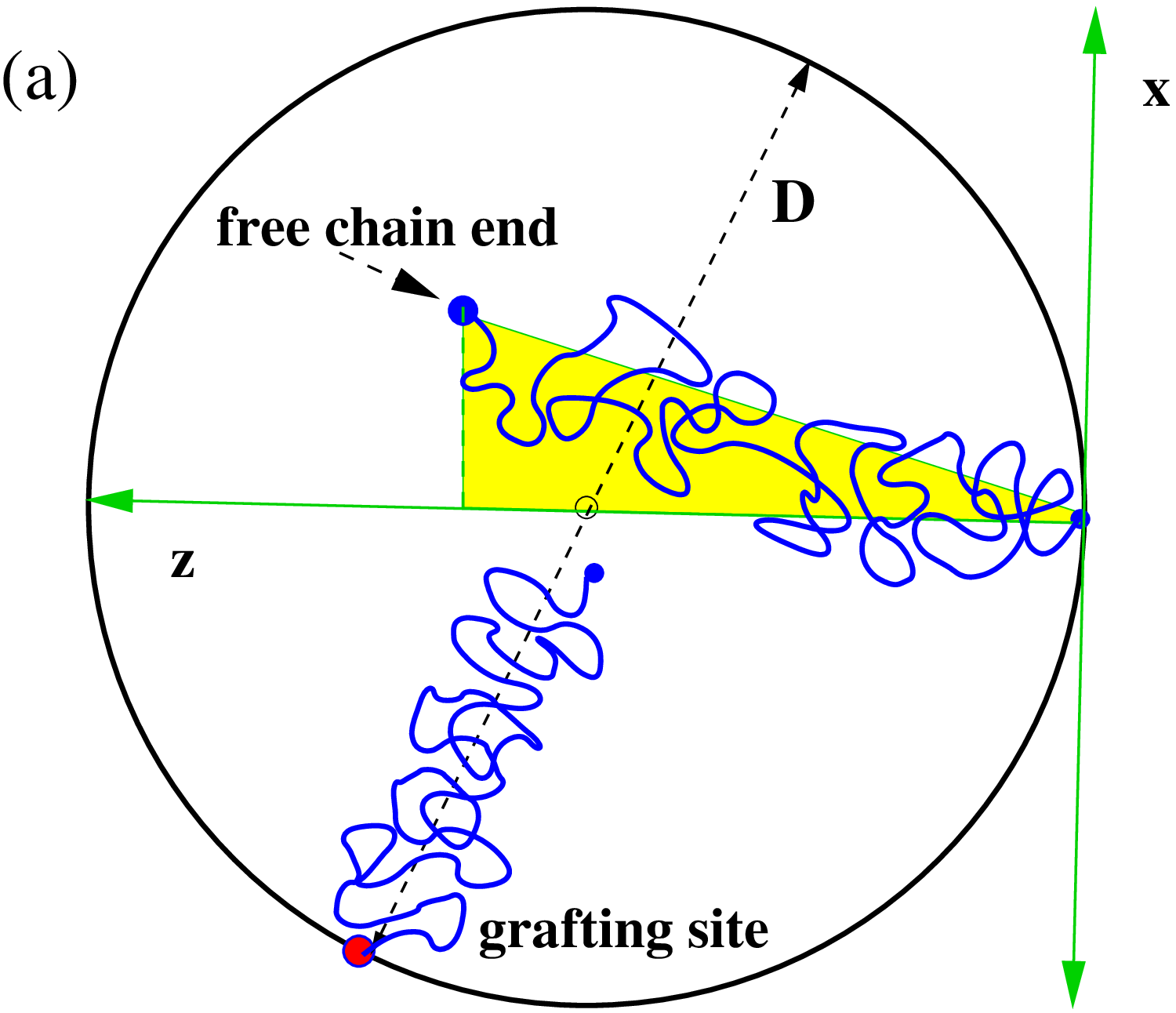}
\hspace*{50pt} \includegraphics[width=2.8in, angle=0]{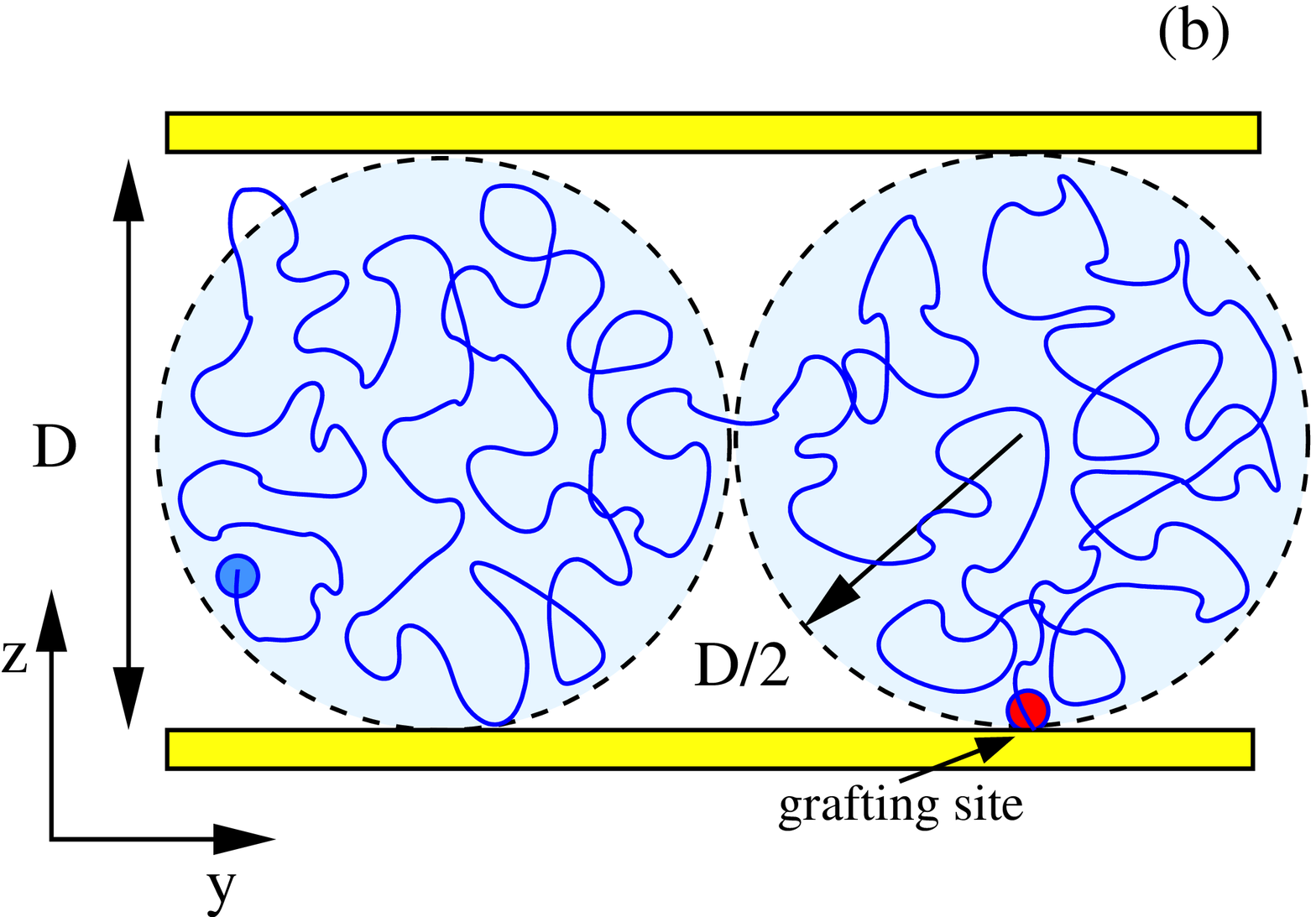}
}
\vspace*{8pt}

\caption{(a) Choice of coordinates in a cylindrical tube of
diameter $D$ used for the analysis of the gyration radius components
of the chains grafted on the interior pore surface. For each
polymer chain a separate coordinate system is chosen with the
origin at the grafting site whereby the $z$-axis is oriented
perpendicular to the grafting surface and points towards the tube axis. 
The $y$-axis is  parallel to the tube axis, and the $x$-axis is
normal to both other axes in the tangential plane at the grafting
site. Note that $z$-coordinates larger than $D/2$ are possible. This coordinate
system is also used to study the distribution $\phi(\vec{r})$ of
the monomers and $\rho(\vec{r})$ of the free chain ends. (b)
Schematic blob picture of a ``cigar'', describing the state of a
chain in a narrow tube in the dilute limit.\label{fig2}}
\end{figure}

\begin{figure}
\centerline{\includegraphics[width=3.7in, angle=0]{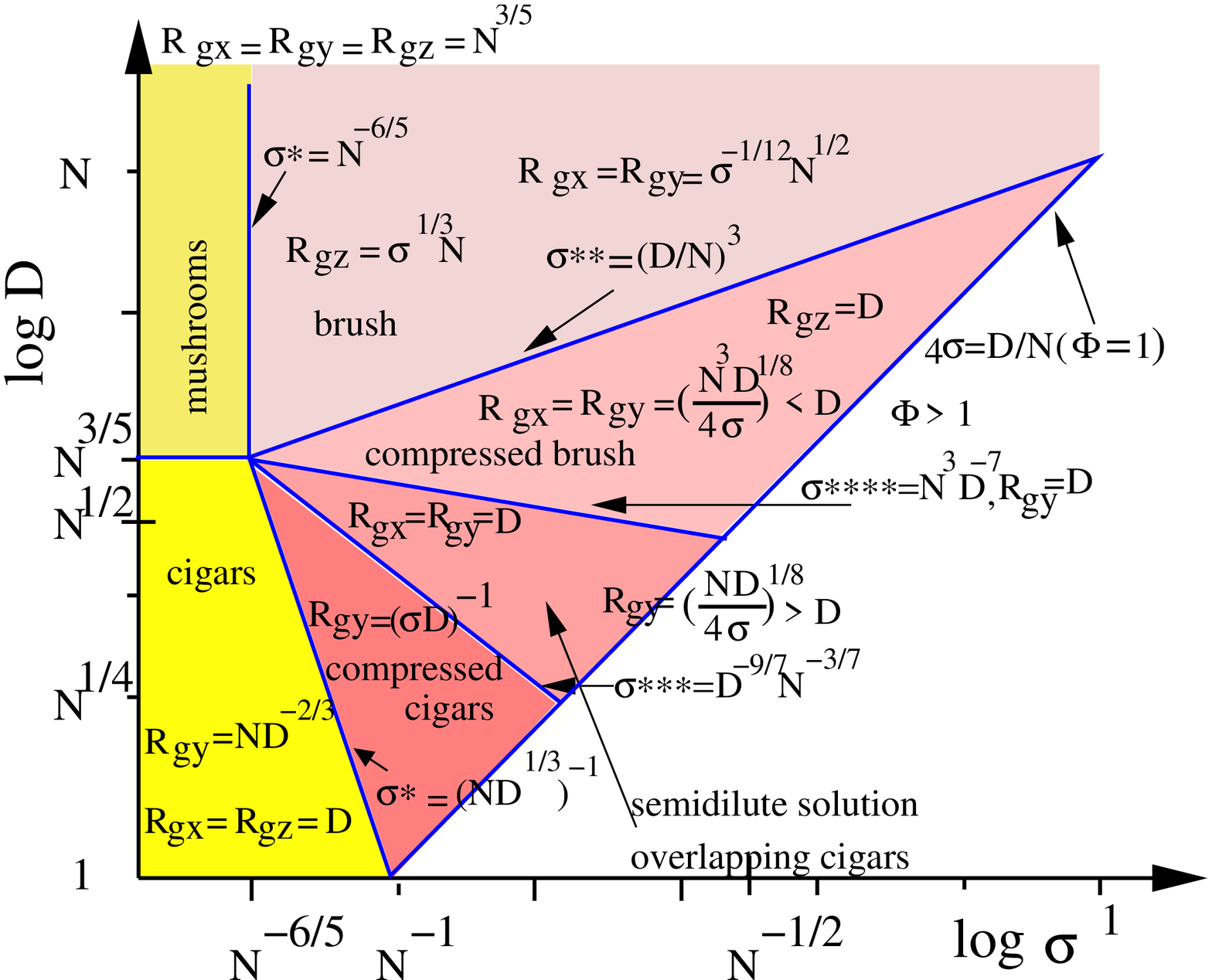}}
\vspace*{8pt}
\caption{\label{fig3} Schematic diagram  for the dimensions of a chain of length
$N$ in terms of tube diameter $D$ and grafting density $\sigma$.
Straight lines indicate the crossovers between the various scaling regimes. 
For {\em wide} tubes ($D >N^{3/5}$) with increasing $\sigma$
only two crossovers are encountered: at $\sigma = \sigma
^*=N^{-6/5}$ from mushrooms to a quasi-flat brush,
and then at $\sigma ^{**}=(D/N)^3$ to a compressed brush. The latter regime
ends at $\sigma = (1/4)D/N$, where the tube is densely filled with monomers 
$(\phi=1)$. Since the line, corresponding to $\phi=1$ (the regime $\phi>1$ 
is unphysical) meets the line $\sigma^{**}$ at $D=N$, for $D>N$ no
compressed brush exists any longer. 
For {\em narrow} tubes, $D<N^{3/5}$, instead of mushrooms one has cigars, 
elongated along the tube axis. Three sub-regimes need to be distinguished:
For moderately narrow tubes, $N^{1/2}<D<N^{3/5}$, three successive
crossovers are distinguished: At $\sigma = \sigma ^*=(ND^{1/3})^{-1}$ a 
crossover from {\em swollen} cigars to {\em compressed} non-overlapping  
cigars occurs. At $\sigma^{***}=D^{-9/7}N^{3/7}$ these cigars start to overlap
strongly whereby for $\sigma^{***}<\sigma <\sigma^{****}=N^3D^{-7}$ their
linear dimension in $y$-direction still exceeds the tube diameter. At
$\sigma^{****}$ the chains behave similar as in a bulk semidilute solution of the
same density and $R_{gy}=D$. For $\sigma >\sigma^{****}$, the regime of the
compressed brush (as described for $D>N^{3/5}$) is entered again.
For narrow tubes, $N^{1/4}<\sigma <N^{1/2}$, the latter regime no
longer exists, however. Finally, for {\em ultranarrow} tubes
($1<D<N^{1/4}$) only a single crossover from swollen to laterally
compressed cigars is possible. 
}
\end{figure}

\begin{figure}
\centerline{\includegraphics[width=4.2in, angle=0]{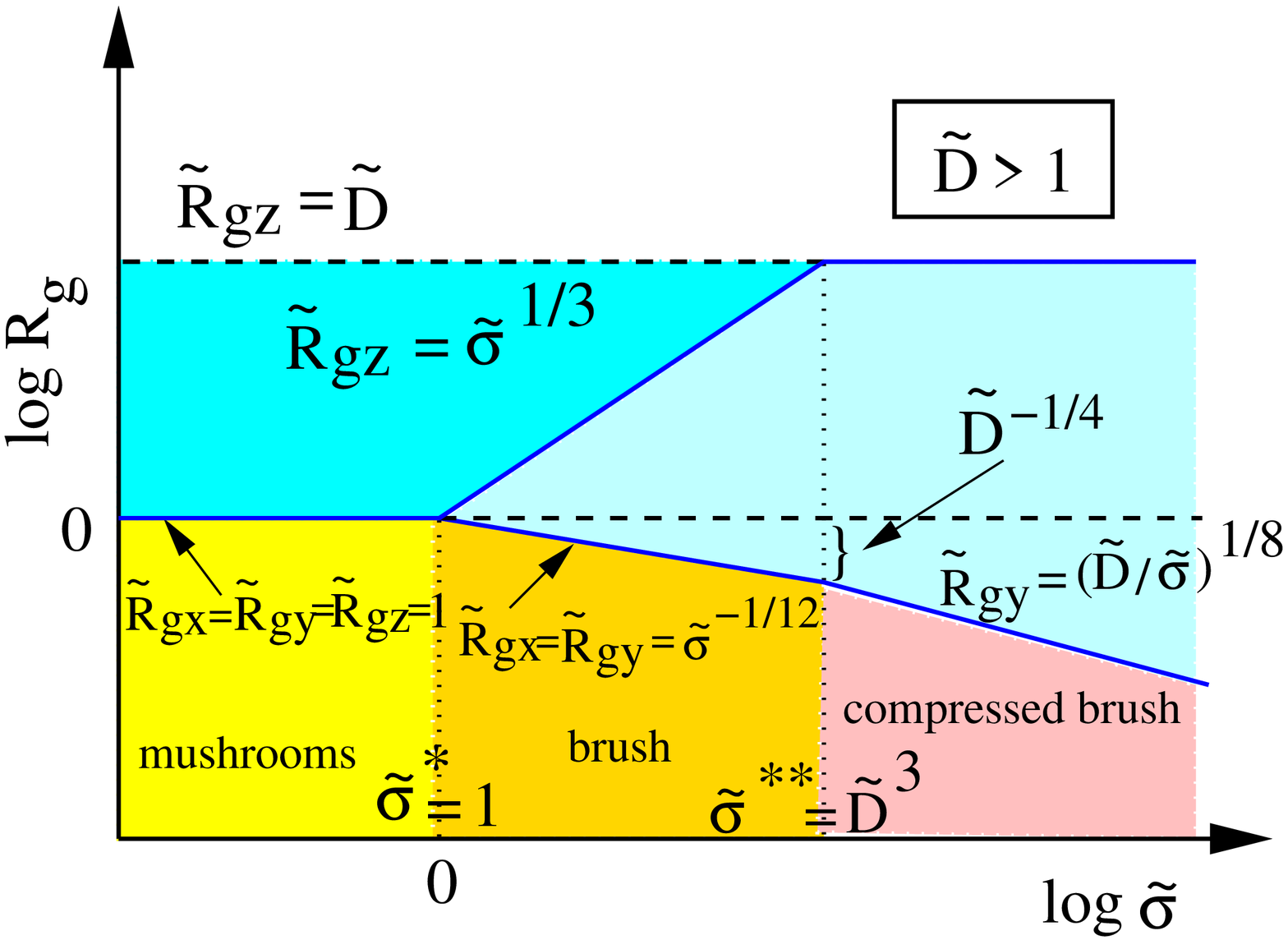}}
\centerline{\includegraphics[width=4.2in, angle=0]{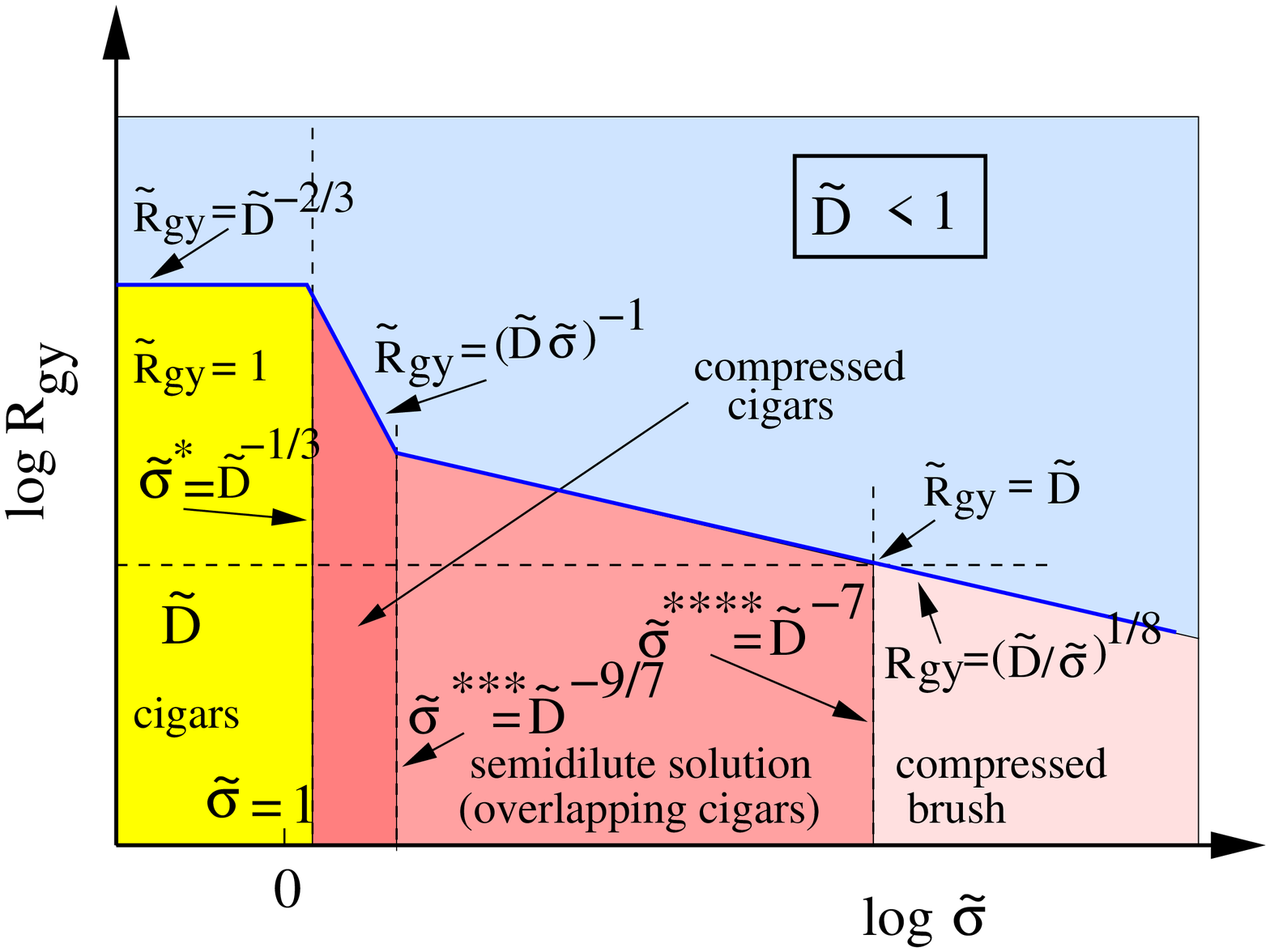}
}
\vspace*{8pt}
\caption{Schematic variation of the scaled linear
dimensions $\tilde{R}_{gy}=R_{gy}/N^{3/5}$,
$\tilde{R}_{gz}=R_{gz}/N^{3/5}$  with the
scaled grafting density $\tilde{\sigma}=\sigma N^{6/5}$ on a
log-log plot. For wide tubes $(\tilde{D}>1$, upper part) and
moderately narrow tubes $(\tilde{D}<1$ but $D>N^{1/2}$, lower part)
there is no longer any dependence on chain length $N$.
Since the crossovers at the various 
grafting densities $\tilde{\sigma}^*,\;\tilde{\sigma} ^{**},
\tilde{\sigma} ^{***}$ and $\tilde{\sigma}^{****}$ are not
sharp, the dashed vertical straight lines should be
understood as centers of crossover regions only, cf. text.\label{fig4}}
\end{figure}

The Monte Carlo simulations yield not only the gyration radius
components of the chains, but also the full density profiles of
the monomers, the distribution function of the chain ends, etc. It
has been found that the chain ends are distributed all over the
cylindrical tube, if the monomer density inside the tube gets
large. While for brushes on planar substrates chain end
distributions still have a broad maximum located roughly near the
brush height, in the cylindrical geometry these distributions are
much broader. It is also remarkable that this distribution spans
the full range of radial distances $(0 <r'<D)$, rather than being
restricted only to the region from the grafting point to the
cylinder axis (Fig.~\ref{fig2}), $0<r'<D/2$. Our findings imply
that theories that approximate the chain end distribution by a
delta function $\delta (r'-h)$ cannot be trusted.
Thus, we identify cases where the brush height in the cylinder 
increases when the cylinder radius is decreased, in contrast to 
predictions of such theories. This means that for smaller $D$ we 
find that the curvature slightly enhances chain
stretching in the direction normal to the grafting substrate, but
this effect is rather small.

Due to technical difficulties of preparing well-equilibrated
configurations of polymer brushes in cylindrical pores, only
rather short chains could be studied, which do not allow a
decisive test of the scaling theory yet. Nevertheless the scaling
theory does provide a useful framework to analyze and discuss the
Monte Carlo results.

In future work, we plan to consider transport through such
polymer-coated nanopipes. We also hope that our study will
stimulate experimental work on such cylindrical brushes. Also a
detailed comparison to self-consistent field treatments would be
valuable.

\underline{Acknowledgement}: Support from the Deutsche
Forschungsgemeinschaft (DFG) under project No 436BUL 113/130 
 is gratefully acknowledged.

\newpage

\clearpage

\clearpage

\clearpage

\clearpage

\begin{figure}
\centerline{\includegraphics[width=4.2in, angle=0]{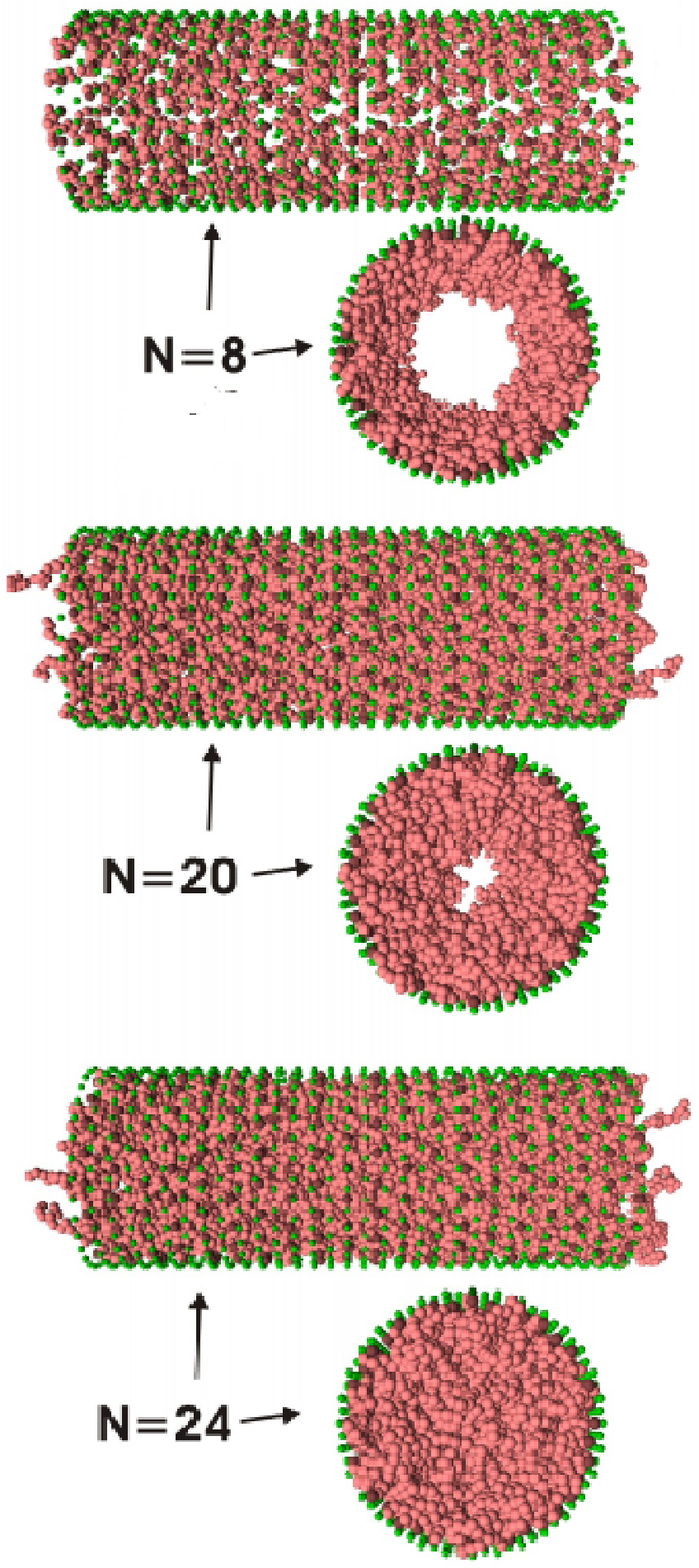}}

\vspace*{8pt}
\caption{Snapshot configurations of the brush in a
tube of diameter $D=30$ and length $L=\pi D$, for three choices of
$N$: $N=8$ (upper part), $N=20$ (middle part) and $N=24$ (lower
part). Both views of the outside and of cross sections are shown.
While for $N=8$ and $N=20$ the center of the tube is typically
still free of monomers, for $N=24$ a uniform dense filling of the
tube is obtained.\label{fig5}}
\end{figure}

\clearpage

\begin{figure}
\centerline{\includegraphics[width=2.7in, angle=-90]{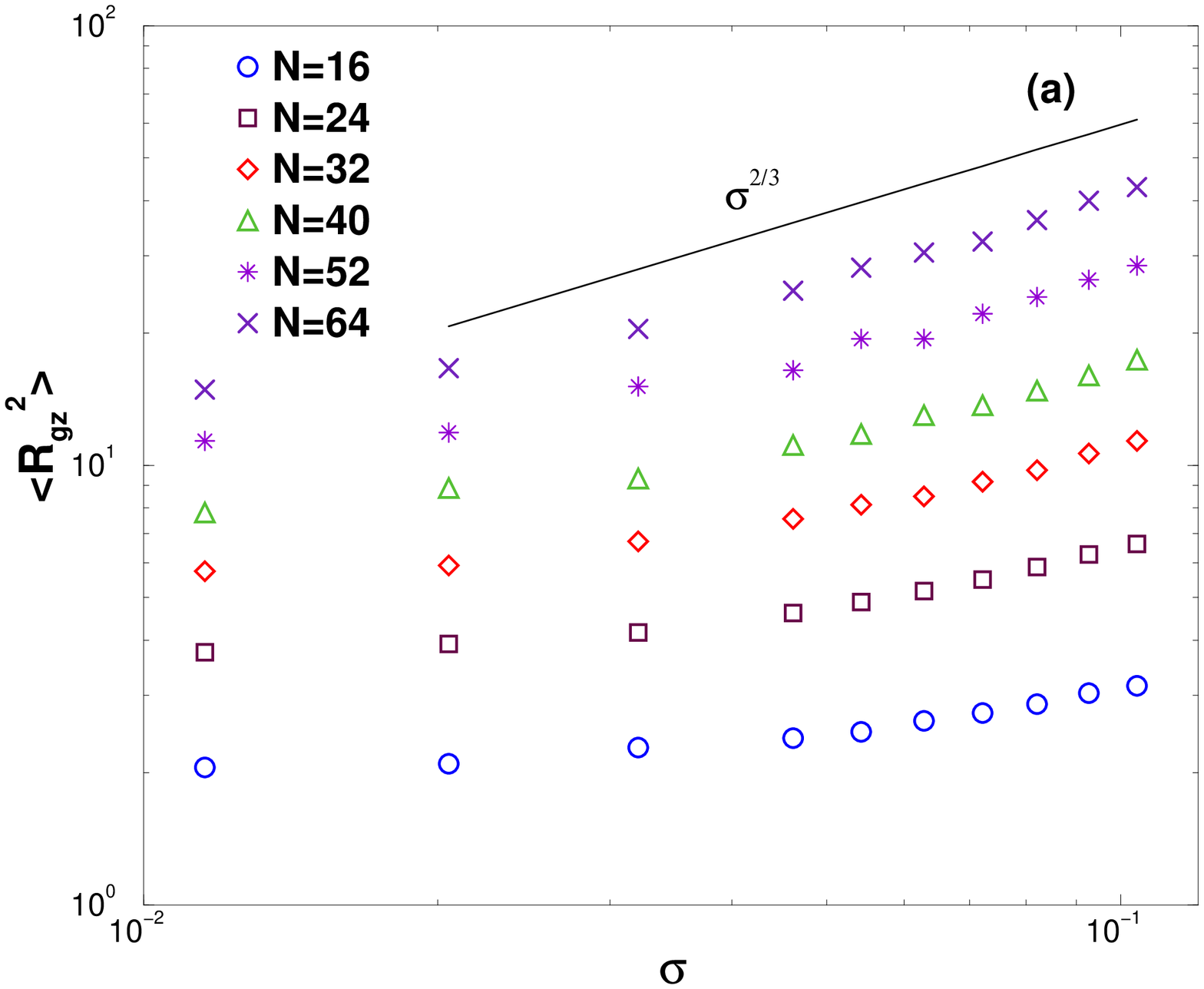}}
\centerline{\includegraphics[width=2.7in, angle=-90]{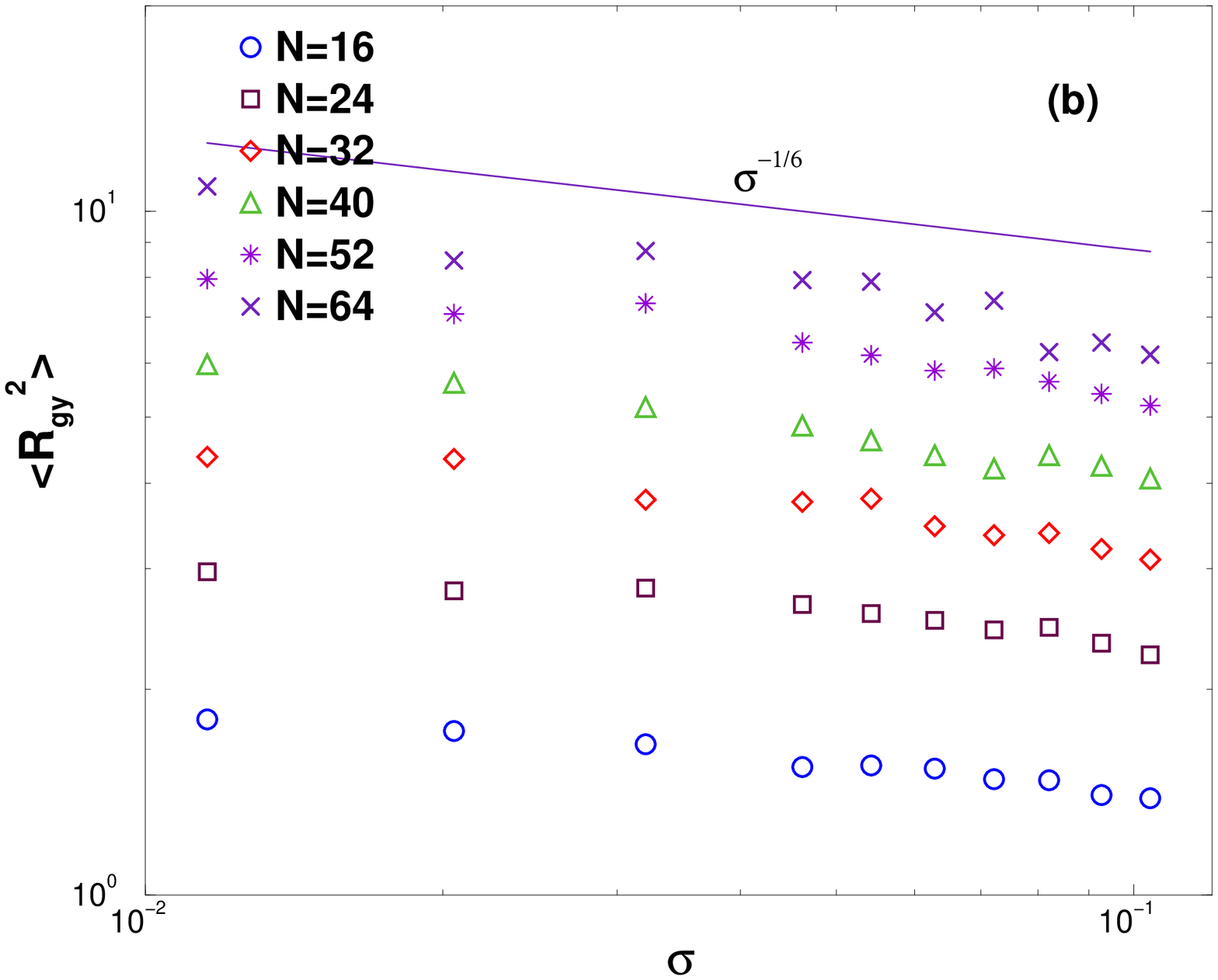}}
\centerline{\includegraphics[width=2.7in, angle=-90]{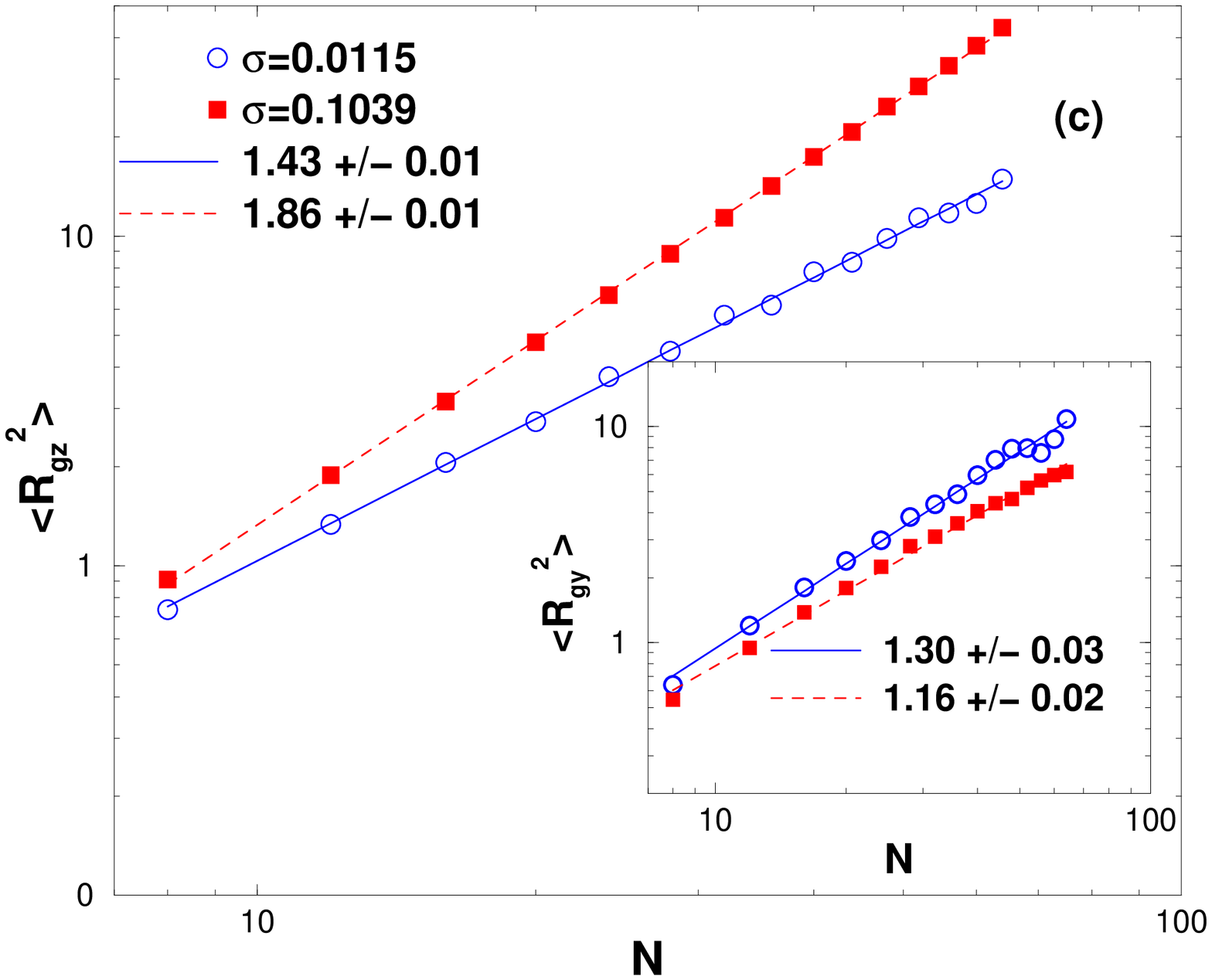}}
\vspace*{8pt}
\caption{Log-log plot of $\langle R^2_{gz}\rangle$
(a) and $\langle R_{gy}^2\rangle $ (b) versus grafting density
$\sigma$, using a perfectly flat substrate. The total number of
chains ranges from $36\; (\sigma = 0.011547)$ to $324\; (\sigma =
0.103926$). Chain lengths $N=16, 24, 32, 40,52$ and $64$ are
included in the plot, as indicated. Straight lines indicate the
theoretical power laws $\langle R_{gz}^2 \rangle \propto
\sigma^{2/3}$ and $\langle R_{gy}^2\rangle \propto \sigma ^{-1/6}$
that apply in the semidilute regime of strongly stretched brushes
under good solvent conditions (Eqs.~(\ref{eq2}), (\ref{eq3})).
Upper part of (b) shows a plot of $\langle R^2_{gy}\rangle
$ vs $N$, while (c) shows a plot of $\langle R^2_{gz}
\rangle $ vs. $N$.\label{fig6}}
\end{figure}

\clearpage

\begin{figure}
\centerline{\includegraphics[width=3.7in, angle=-90]{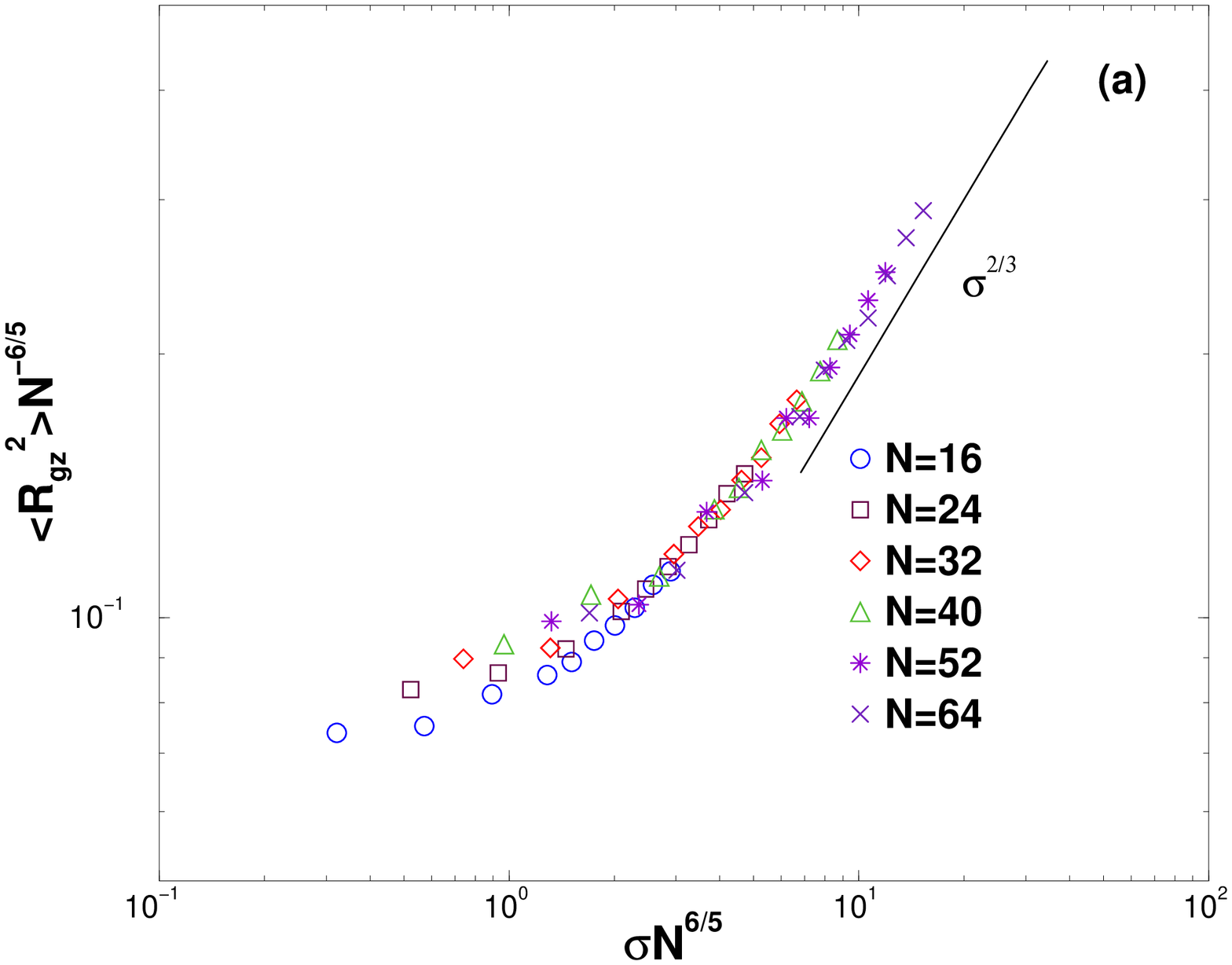}}
\centerline{\includegraphics[width=3.7in, angle=-90]{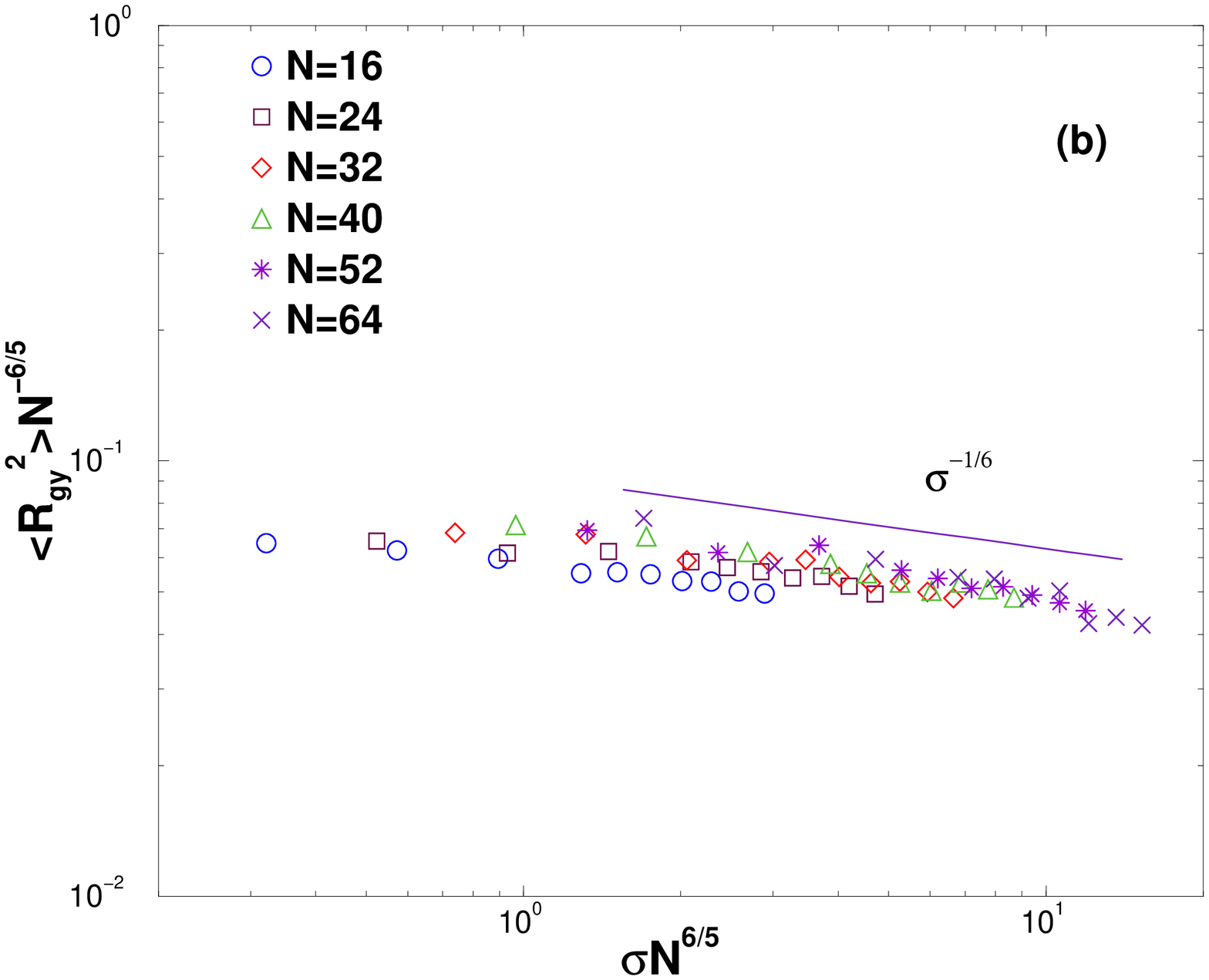}}
\vspace*{8pt}
\caption{Scaling plots where $\tilde{R}_{gz}^2 =
\langle R^2_{gz} \rangle /N^{6/5}$ (a) and $\tilde{R}^2_{gy} =
\langle R^2_{gy} \rangle/N^{6/5}$ (b) are shown as functions of
the scaled grafting density $\sigma N^{6/5}$, in a log-log
representation. Same data as shown in Fig.~\ref{fig6} are used,
and again the theoretical power laws $\tilde{R}_{gz}^2 \propto
\tilde{\sigma}^{2/3}$ and $\tilde{R}_{gy}^2 \propto
\tilde{\sigma}^{-1/6}$ are indicated by straight
lines.\label{fig7}}
\end{figure}

\clearpage

\begin{figure}
\centerline{\includegraphics[width=3.7in,
angle=-90]{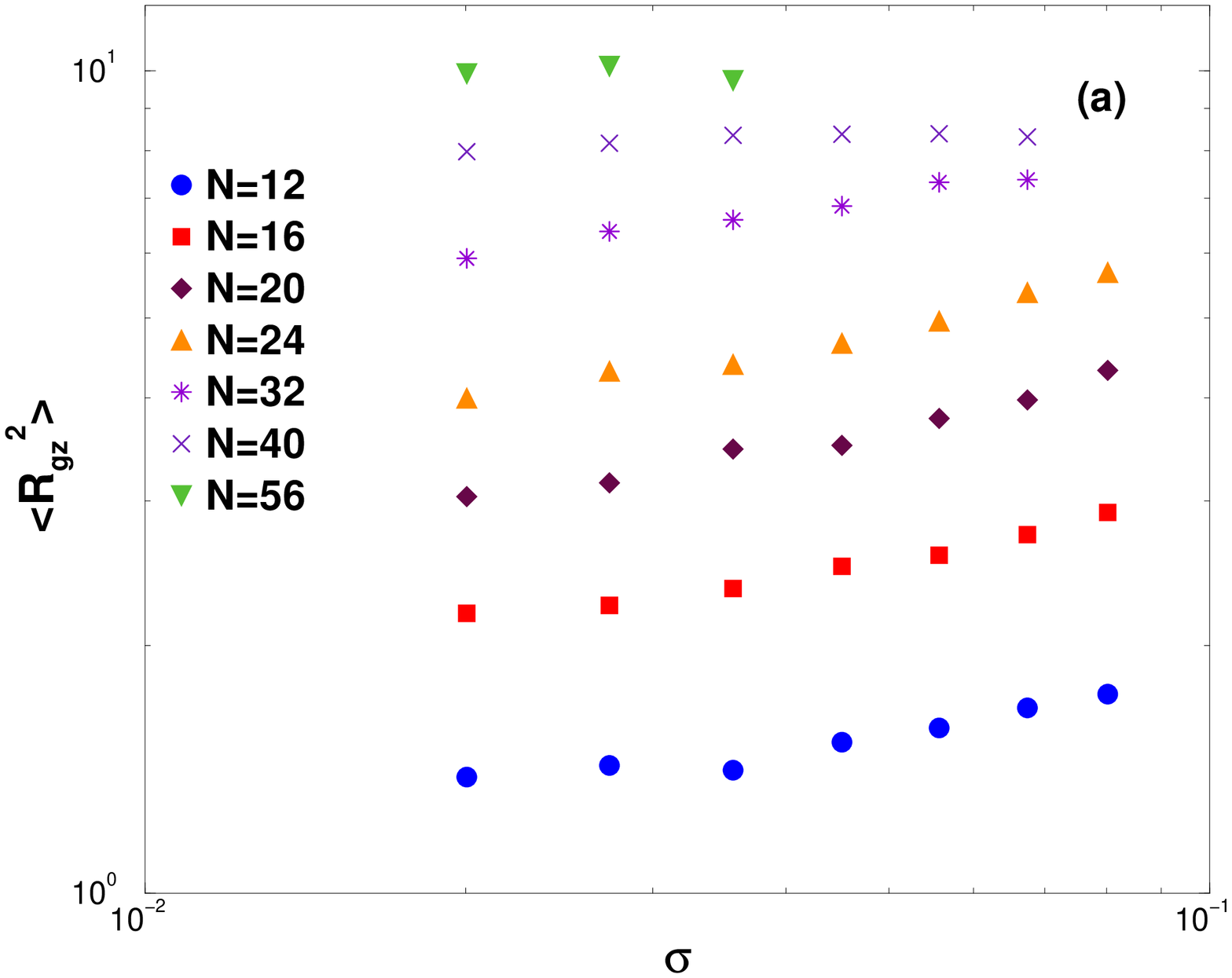}}
\centerline{\includegraphics[width=3.7in,
angle=-90]{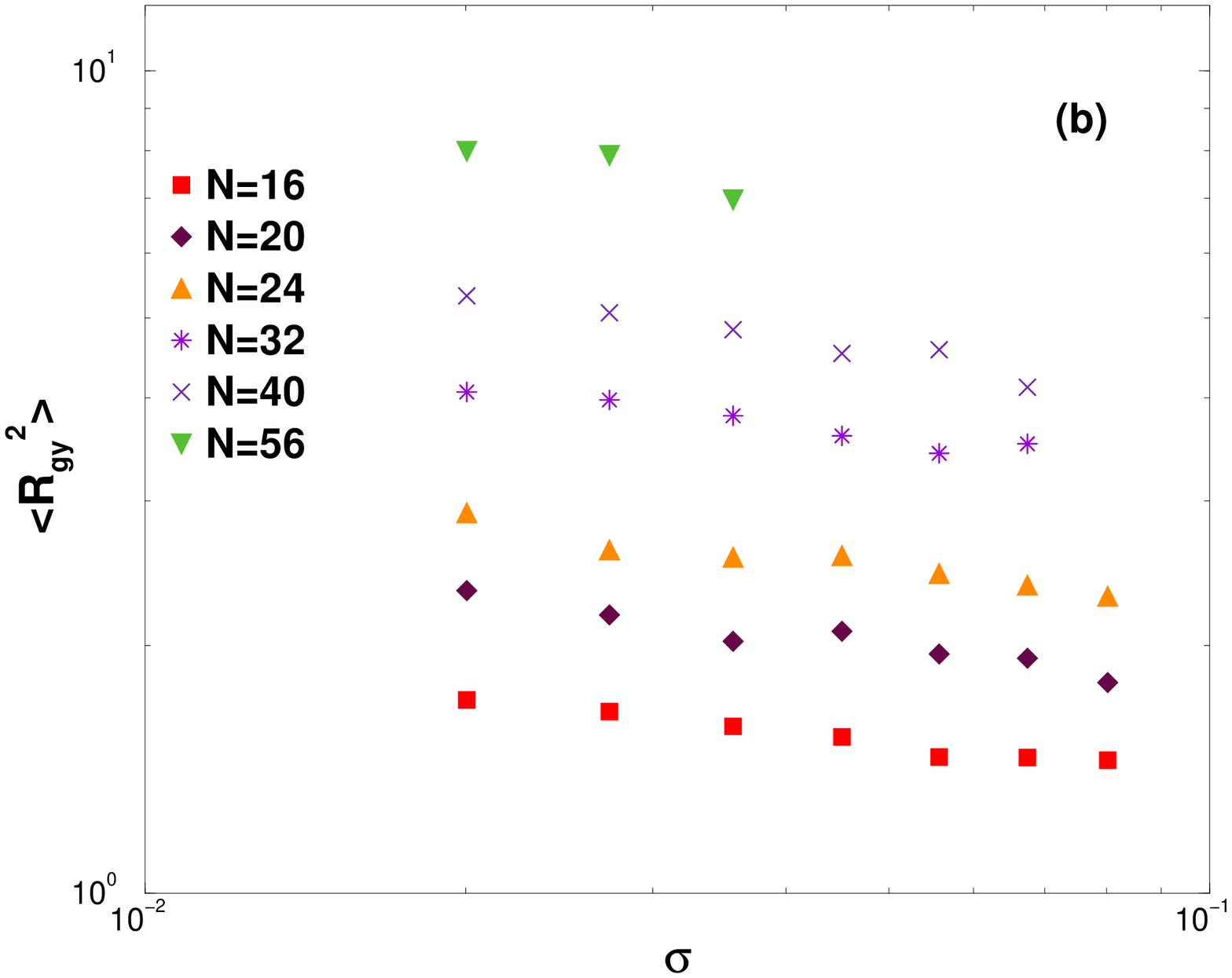}} \vspace*{8pt} \caption{Log-log plot of
$\langle R^2_{gz}\rangle$ (a) and $\langle R^2_{gy} \rangle$ (b)
versus grafting density $\sigma$, for a brush in a tube of
diameter $D=30$. The total number of chains ranged from $144
\;(\sigma = 0.02$) to $576\; (\sigma =0.08$). For $\sigma
\leq 0.0356$ chain lengths $N=12, 16, 20, 24, 32, 40$ and $56$ are
included, while for larger $\sigma$ some of the larger chain
lengths needed to be omitted since melt densities were reached
inside the pore (recall Fig.~\ref{fig3}: the unphysical regime
cannot be entered).\label{fig8}}
\end{figure}

\clearpage

\begin{figure}
\centerline{\includegraphics[width=3.7in,
angle=-90]{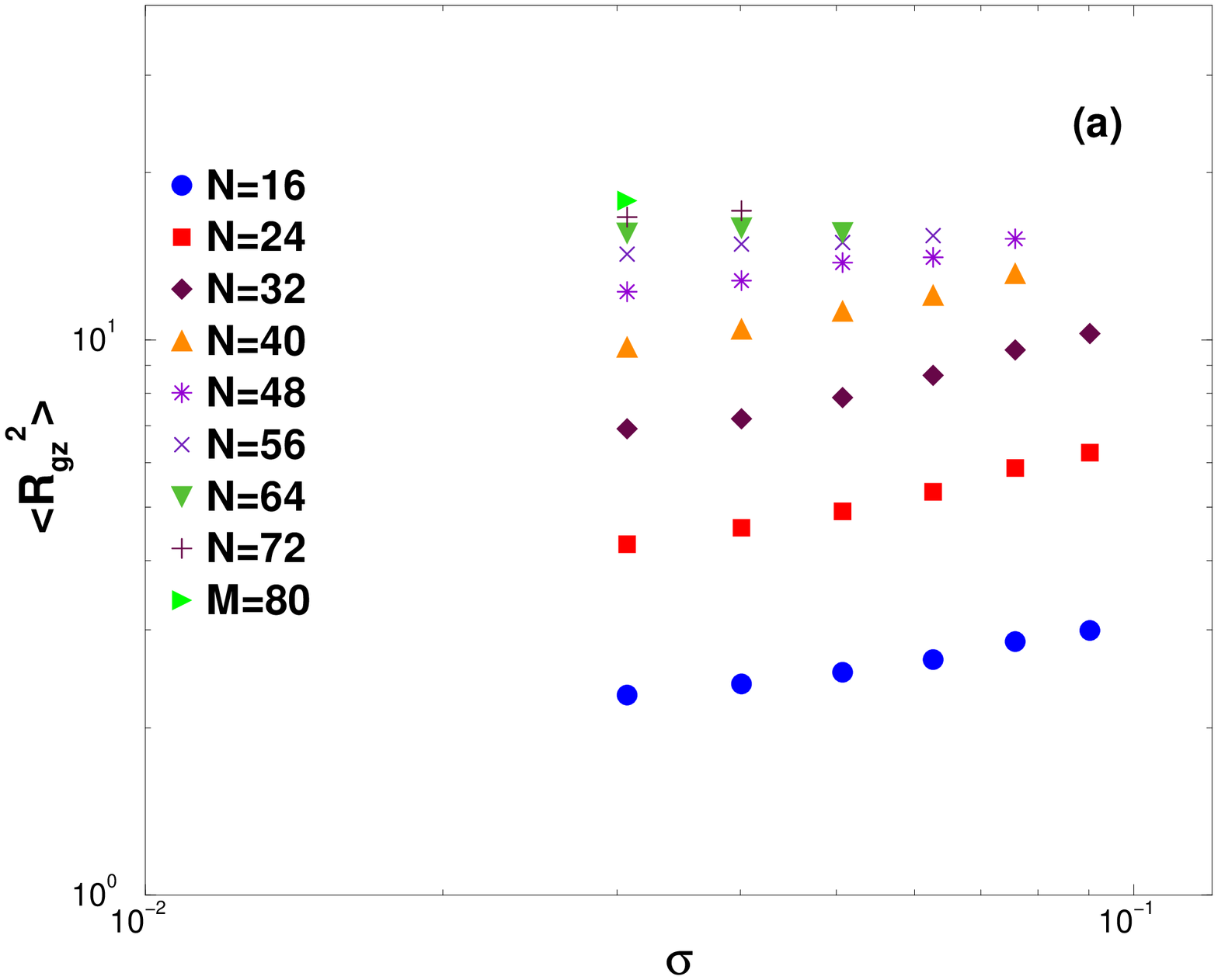}}
\centerline{\includegraphics[width=3.7in,
angle=-90]{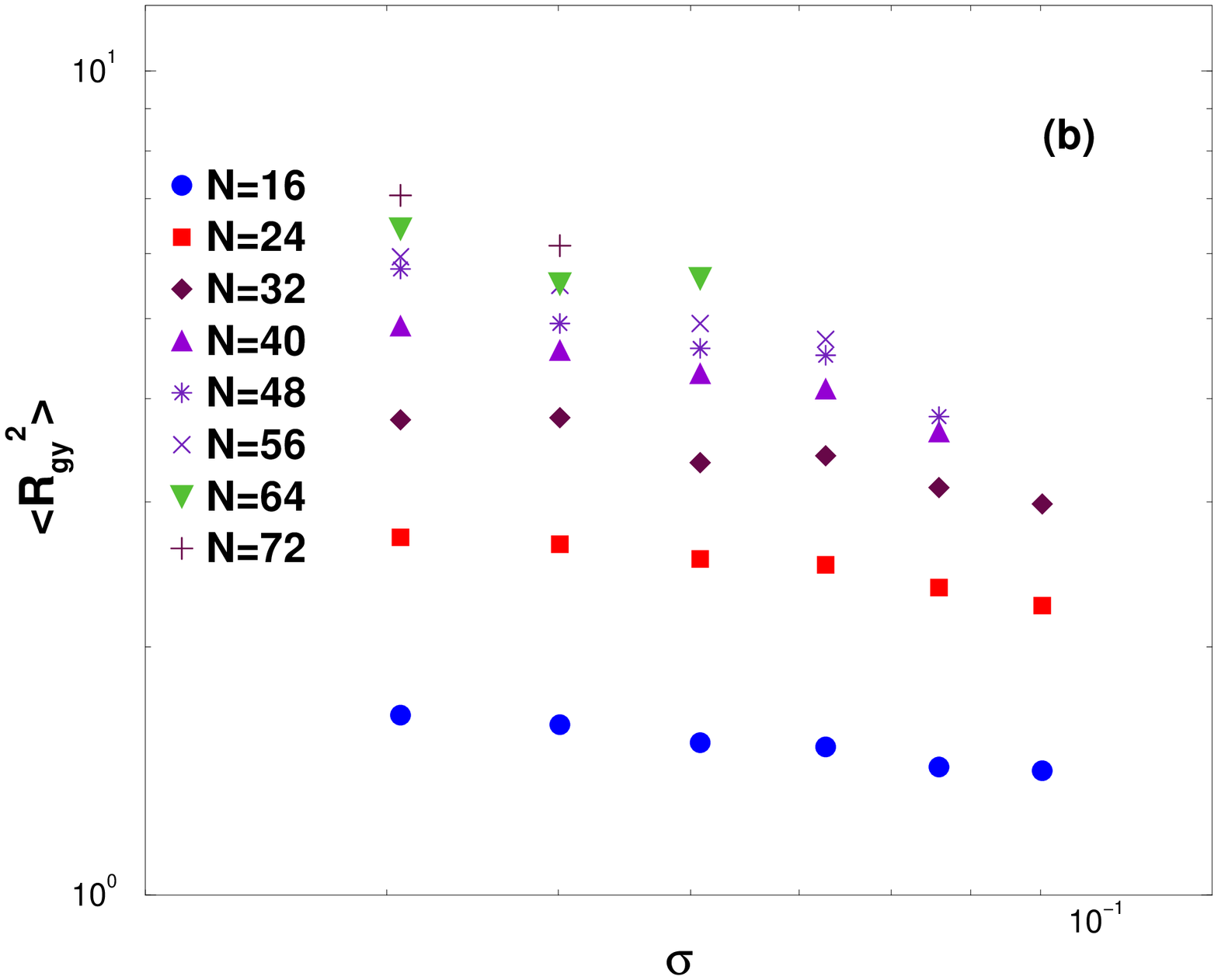}} \vspace*{8pt} \caption{Same as
Fig.~\ref{fig8}, but for $D=42$. The total number of chains ranges
from $294\; (\sigma = 0.031$) to $864\; (\sigma = 0.09$). Chain
lengths up to $N=80$ are included for the smallest and up to
$N=32$ for the largest grafting density.\label{fig9}}
\end{figure}

\clearpage

\begin{figure}
\centerline{\includegraphics[width=2.7in,
angle=-90]{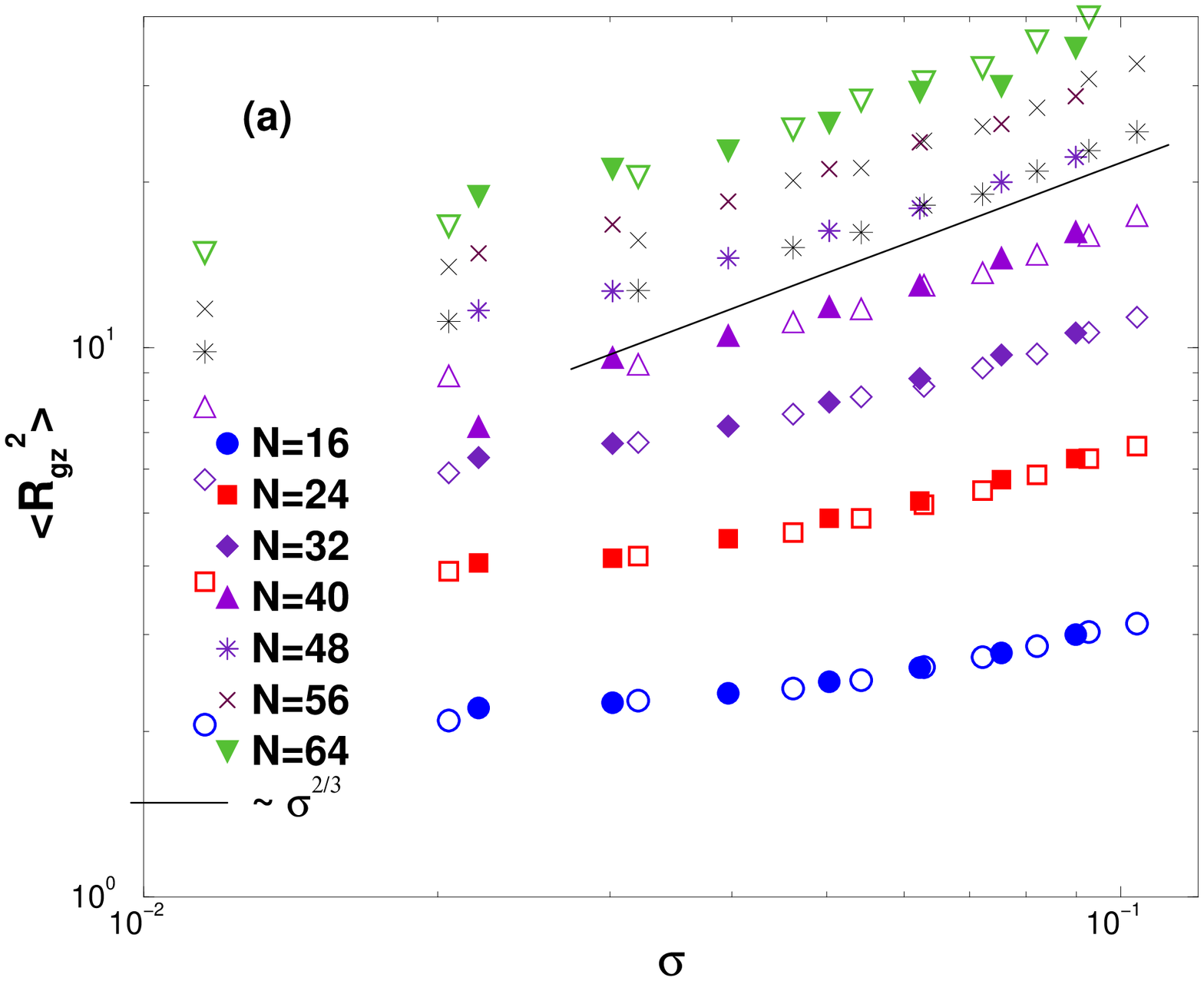}}
\centerline{\includegraphics[width=2.7in,
angle=-90]{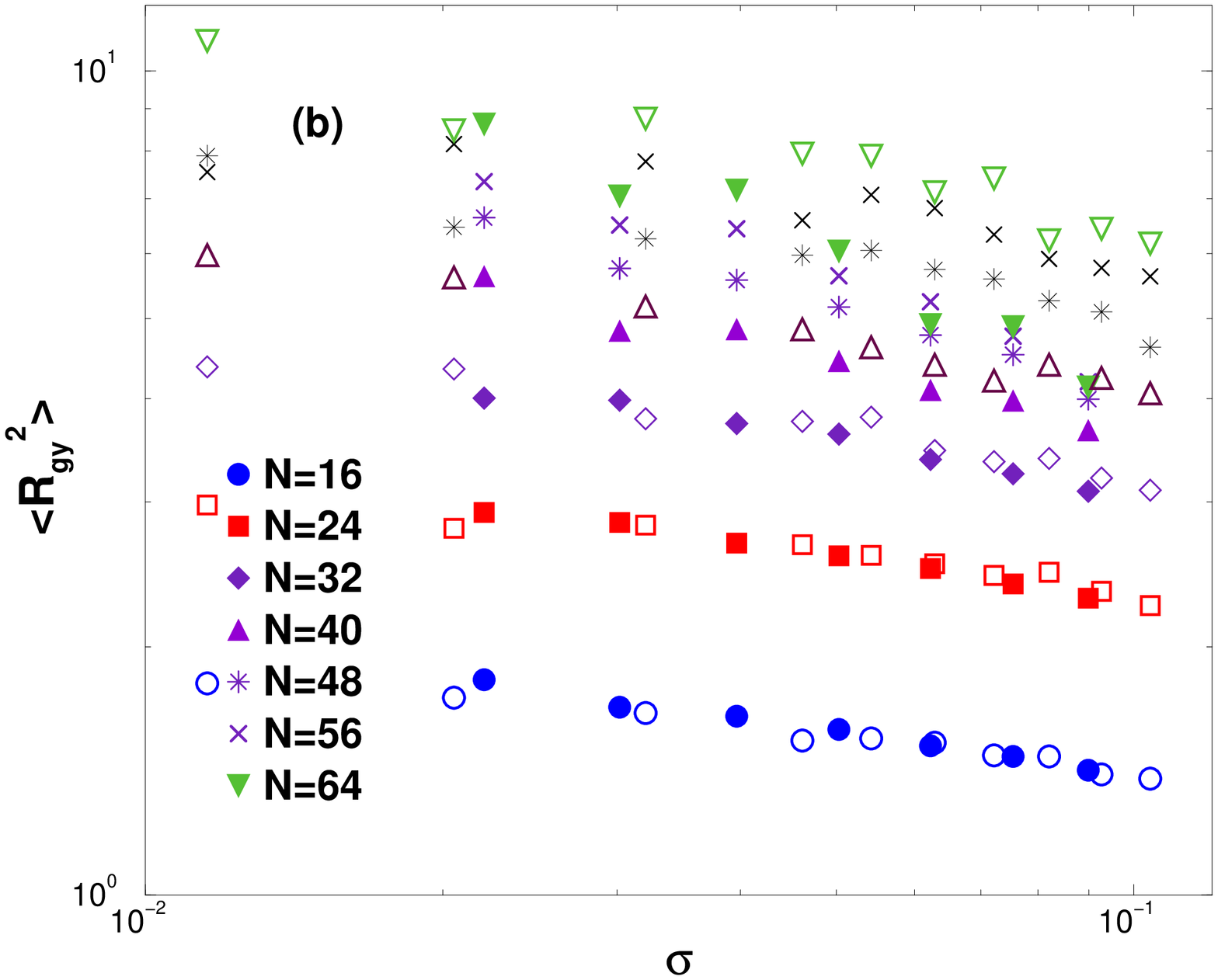}}
\centerline{\includegraphics[width=2.7in,
angle=-90]{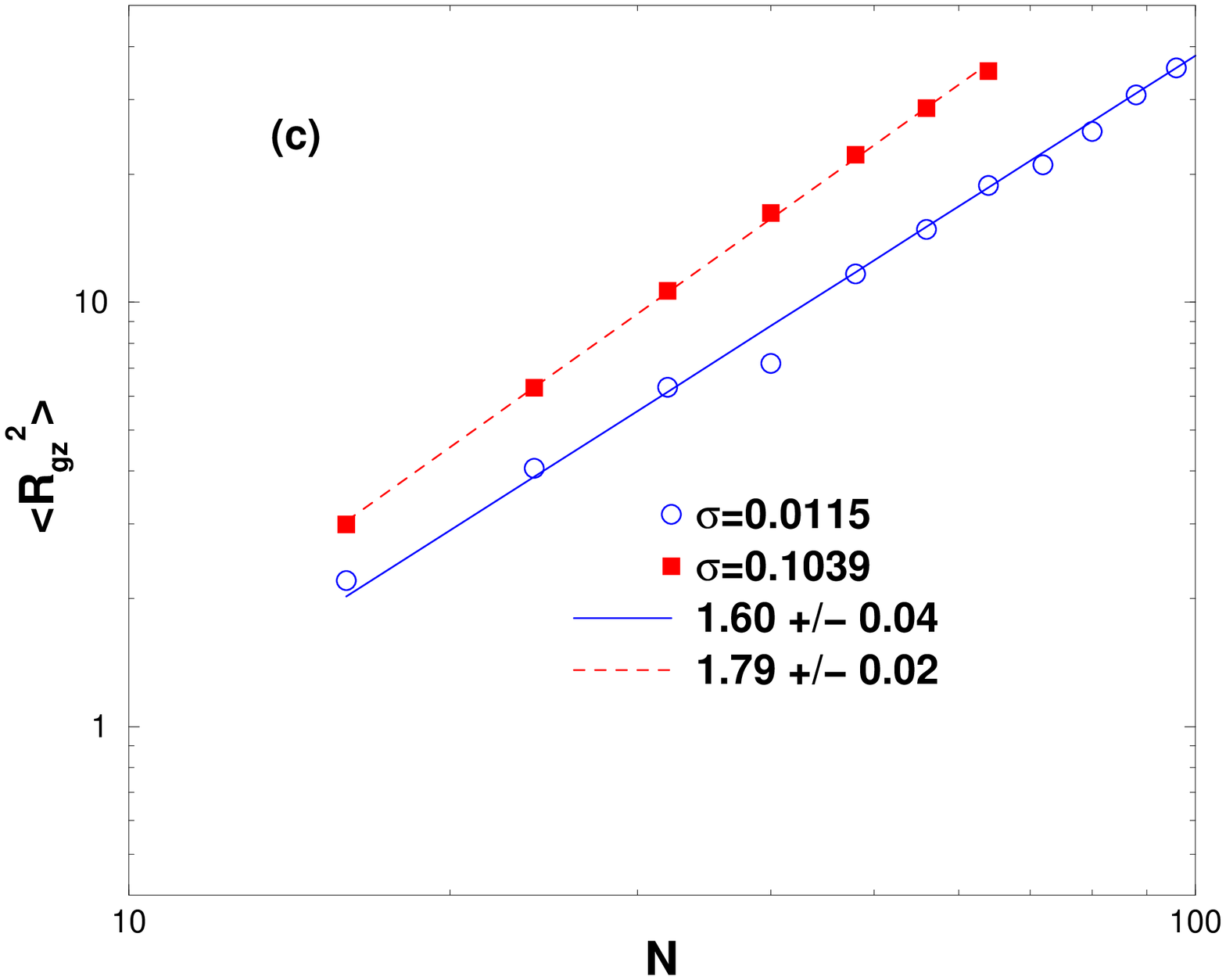}} \vspace*{8pt}
\caption{Same as
Fig.~\ref{fig8}, but for $D=62$. The total number of chains ranges
from $280\; (\sigma = 0.022$) to $1144\;(\sigma = 0.09$).  Part
(c) shows a log-log plot of $\langle R^2_{gz}\rangle $ vs. $N$ for
two grafting densities. In parts (a), (b) we have included
corresponding data for the flat brush on $\langle R^2_{gz} \rangle
$ (empty symbols) to show that at least for $N=32$ and $N=40$ the 
chains for the brush
in the tube are slightly more stretched than for the flat
brush.\label{fig10}}
\end{figure}

\clearpage

\begin{figure}
\centerline{\includegraphics[width=3.7in,
angle=-90]{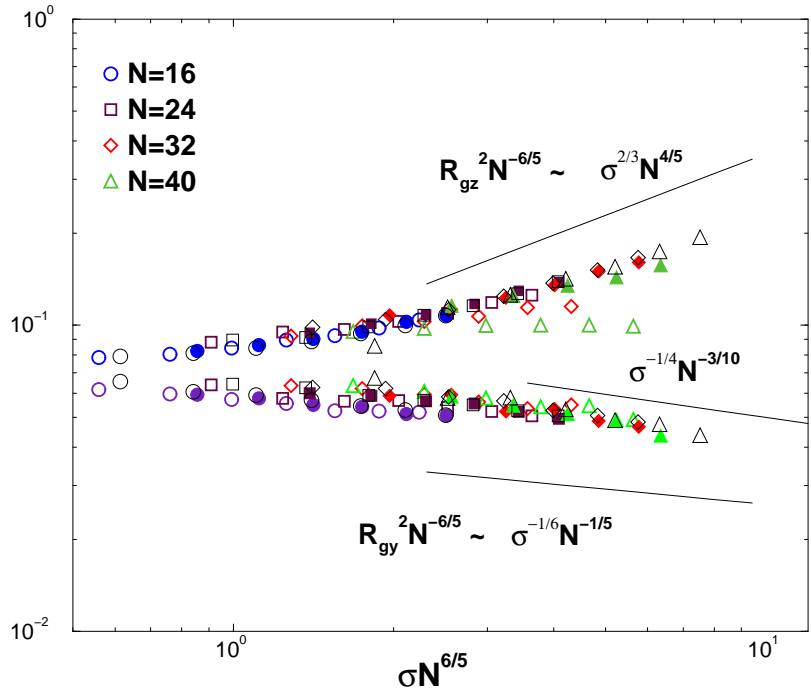}} \vspace*{8pt} \caption{Scaling
representation where $\tilde{R}_{gz}^2$ (upper part) and
$\tilde{R}^2_{gy}$ (lower part) is plotted versus $\tilde{\sigma}$
in double-logarithmic form, including data for various $N$ and $D$.
Open symbols: (thick small) $D=30$, (thin big) $D=62$. Full
symbols: $D=42$. Predicted power laws for the brush regime and the
regime of the compressed brush are shown by straight
lines.\label{fig11}}
\end{figure}

\clearpage

\begin{figure}
\centerline{\includegraphics[width=2.7in, angle=-90]{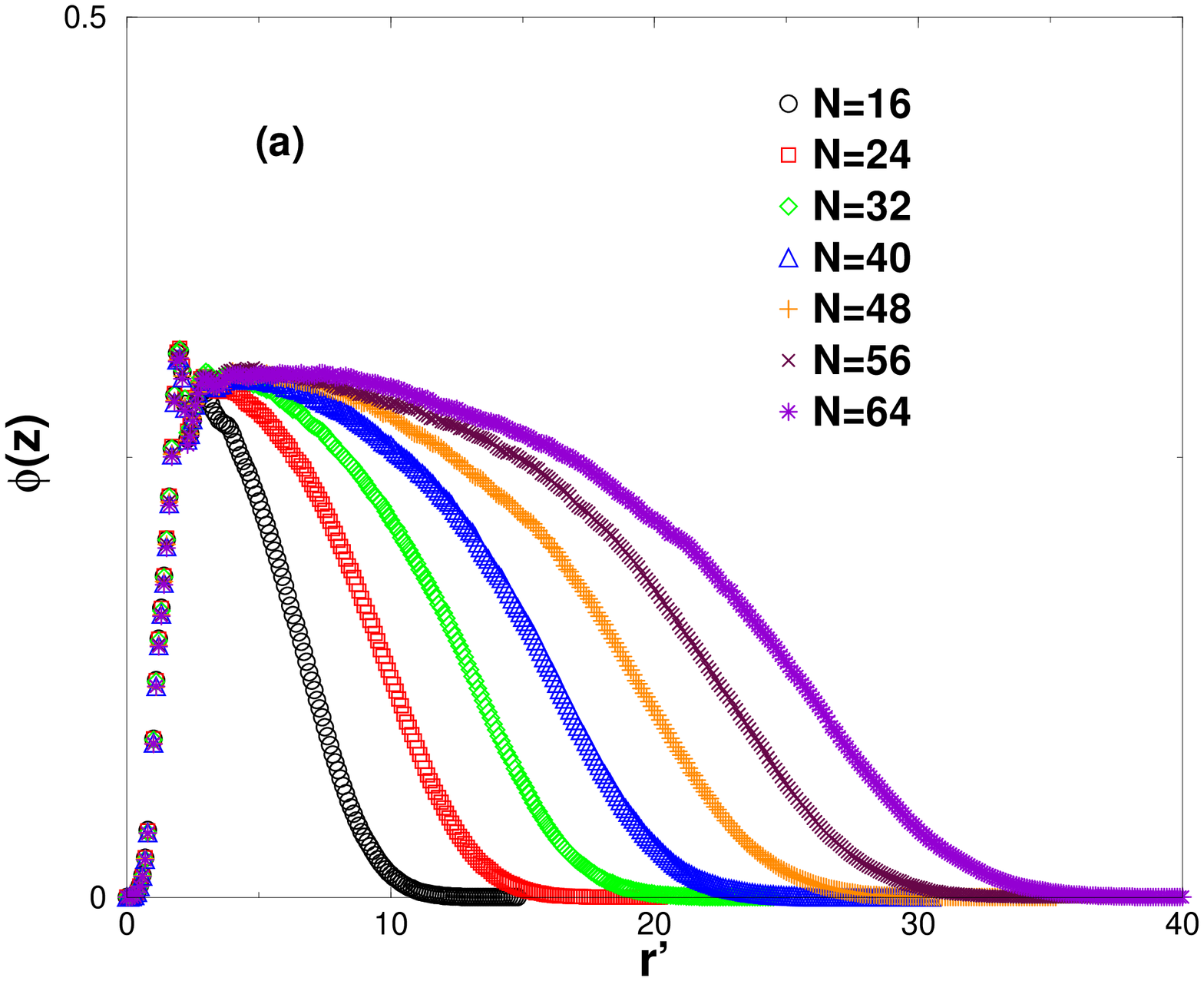}}
\centerline{\includegraphics[width=2.7in, angle=-90]{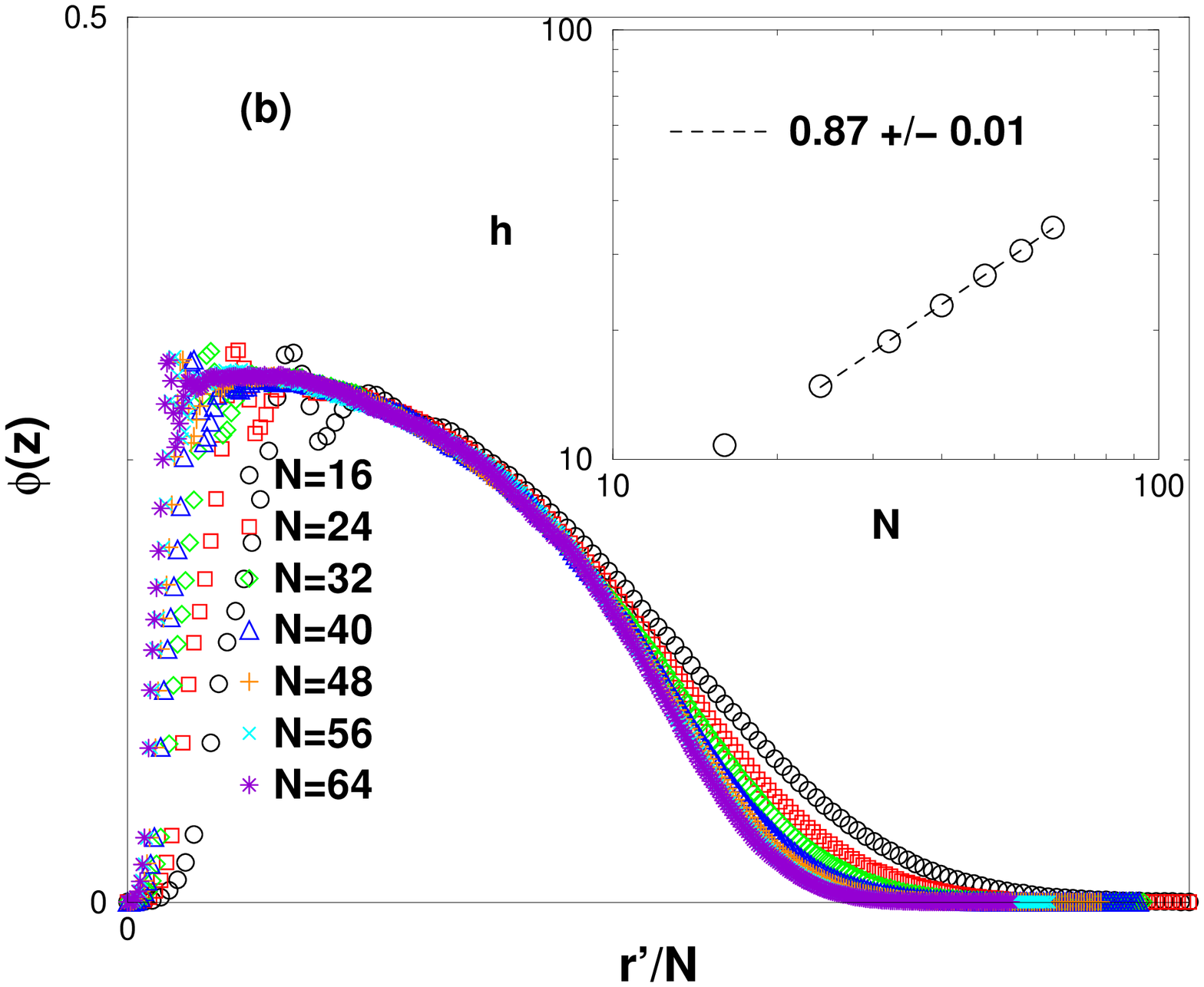}}
\centerline{\includegraphics[width=2.7in, angle=-90]{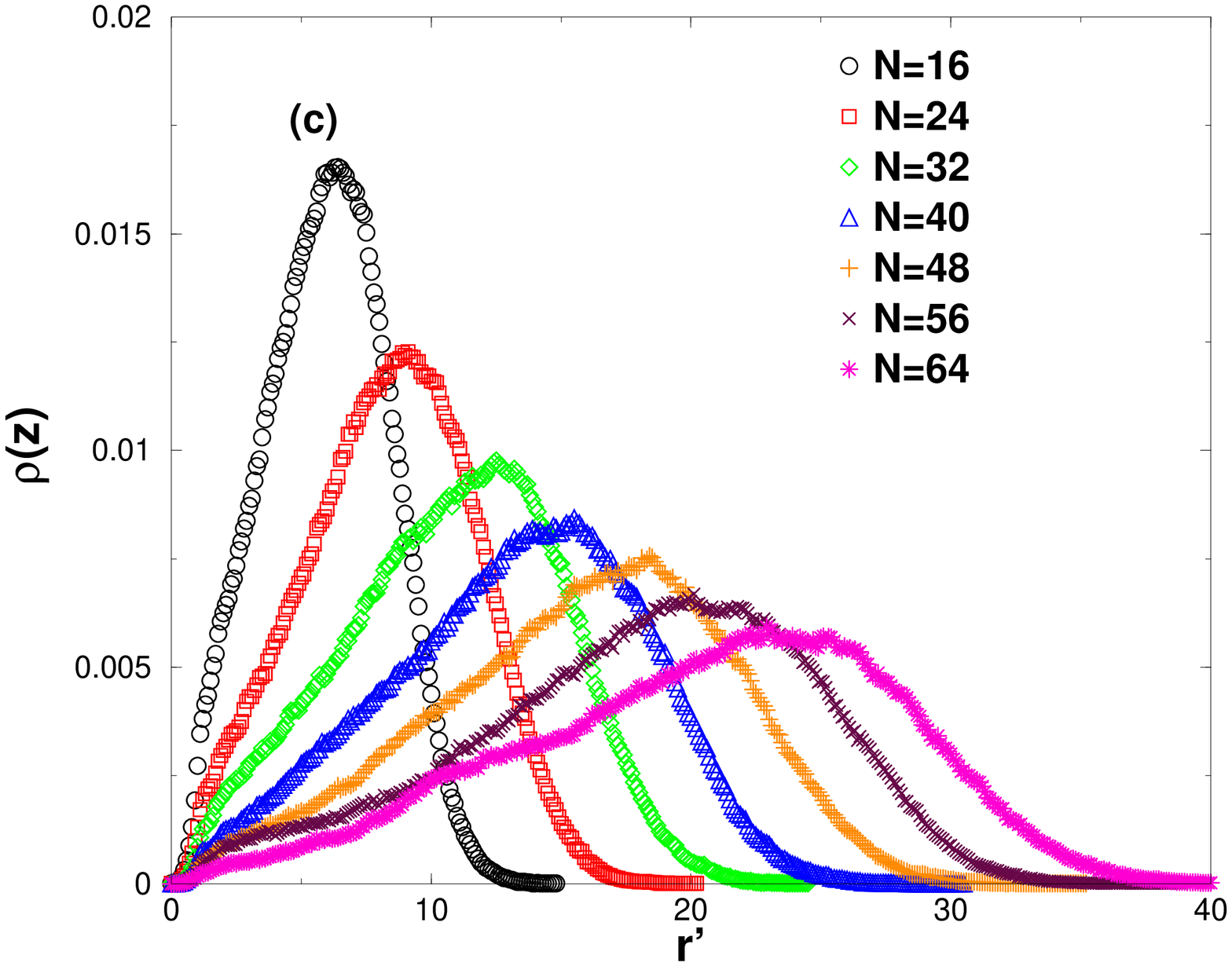}}
\vspace*{8pt}
\caption{Monomer density profile $\phi(r')$ of a
polymer brush on a flat substrate (a) and plotted versus
the rescaled distance $r'/N$ (b). All data refer to a
grafting density $\sigma = 0.1$ ($N_{ch}=324$ chains 
grafted to an area $A= 3117.6$). Seven chain lengths $N$ are
included in the figure, as indicated. The insert in the lower part
shows a log-log plot of the brush height $h$ versus chain length
$N$. The distribution of free chain ends is shown in (c).\label{fig12}}
\end{figure}

\clearpage

\begin{figure}
\centerline{\includegraphics[width=2.7in,
angle=-90]{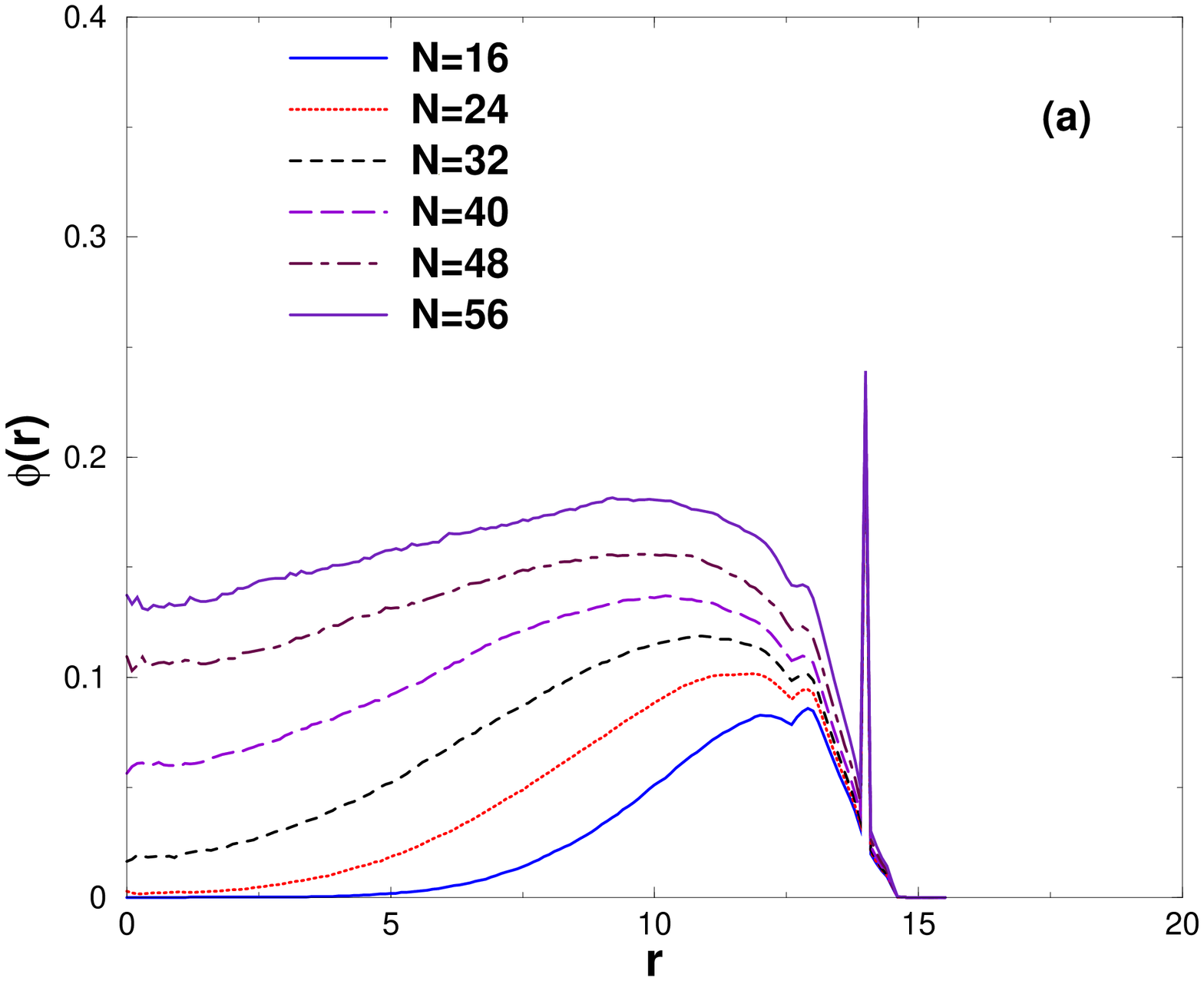}}
\centerline{\includegraphics[width=2.7in,
angle=-90]{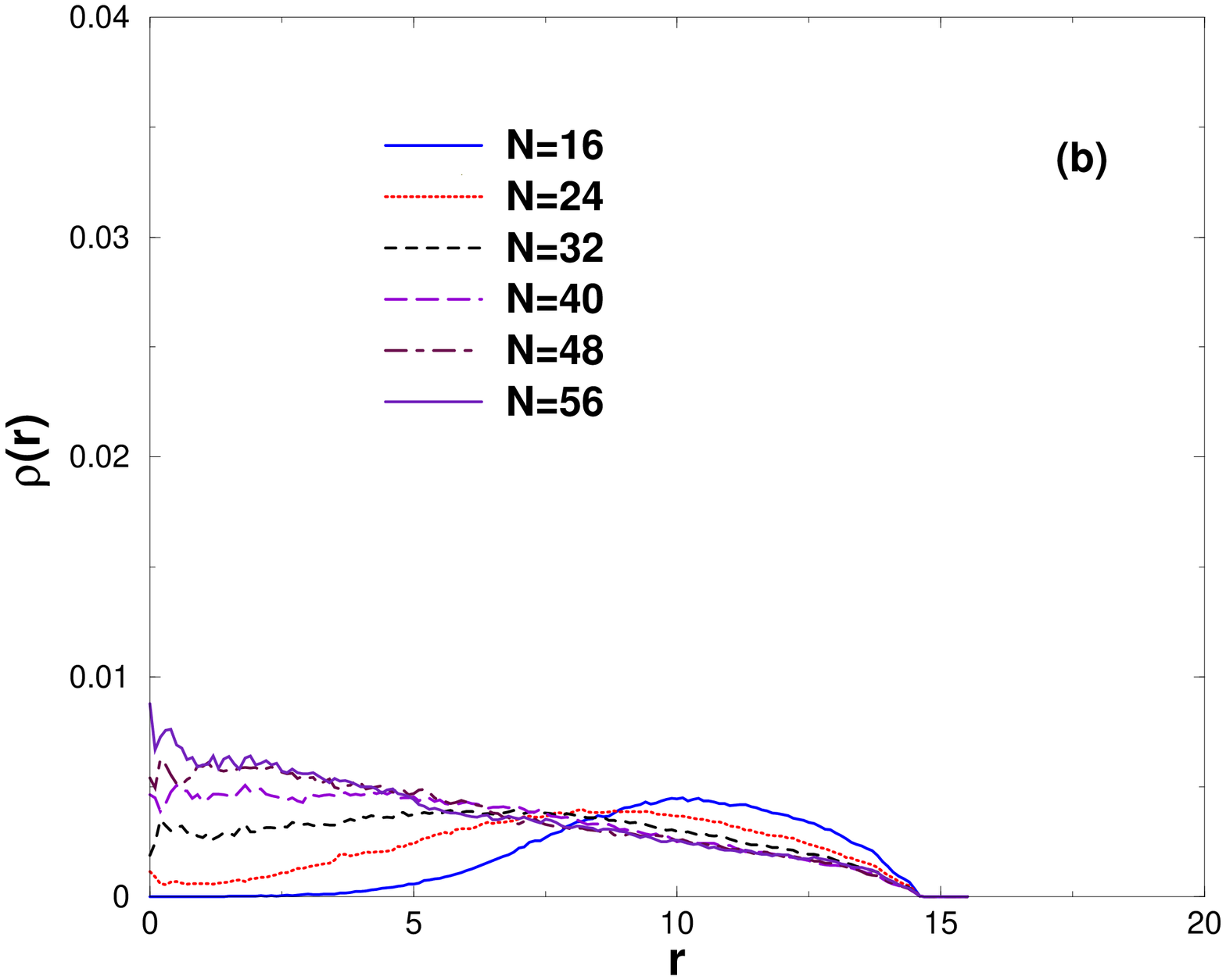}} \vspace*{8pt} \caption{Radial density
distribution function $\phi (r)$ of the monomers as a
function of the distance $r$ from the cylinder axis, for a tube of
radius $R=15$, at a grafting density $\sigma = 0.02$
($N_{ch}=144$ chains), upper part, and the end monomer density
distribution, lower part. Remember that at $r=R-1=14$ the first
monomer of each grafted chain is located, while the atoms forming
the wall are located at $r=R=15$. \label{fig13}}
\end{figure}

\clearpage

\begin{figure}
\centerline{\includegraphics[width=2.7in, angle=-90]{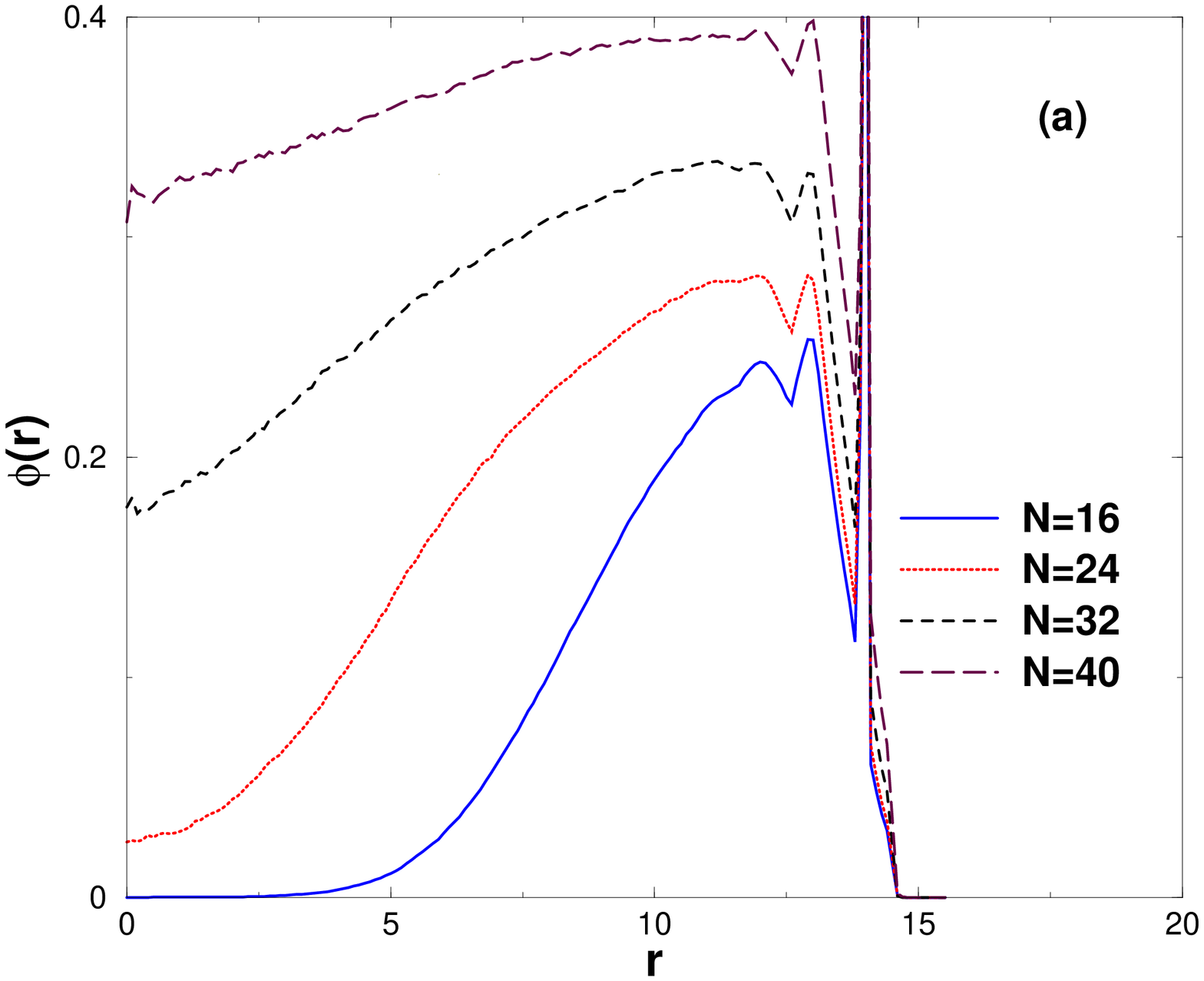}}
\centerline{\includegraphics[width=2.7in, angle=-90]{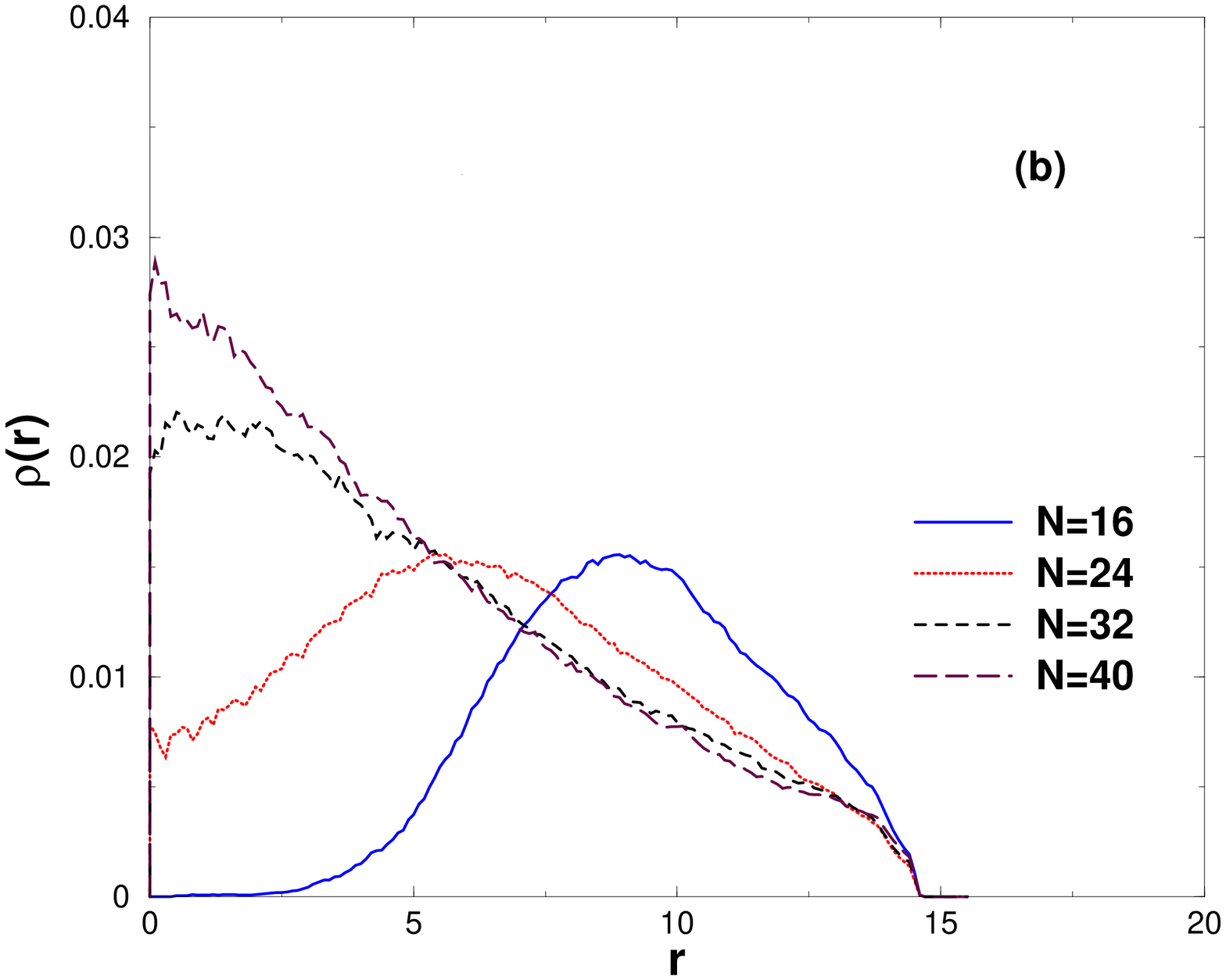}}
\vspace*{8pt}
\caption{Same as Fig.~\ref{fig14}, but for $\sigma =
0.0674$ ($N_{ch}=484$) \label{fig14}}
\end{figure}

\clearpage

\begin{figure}
\centerline{\includegraphics[width=2.7in,
angle=-90]{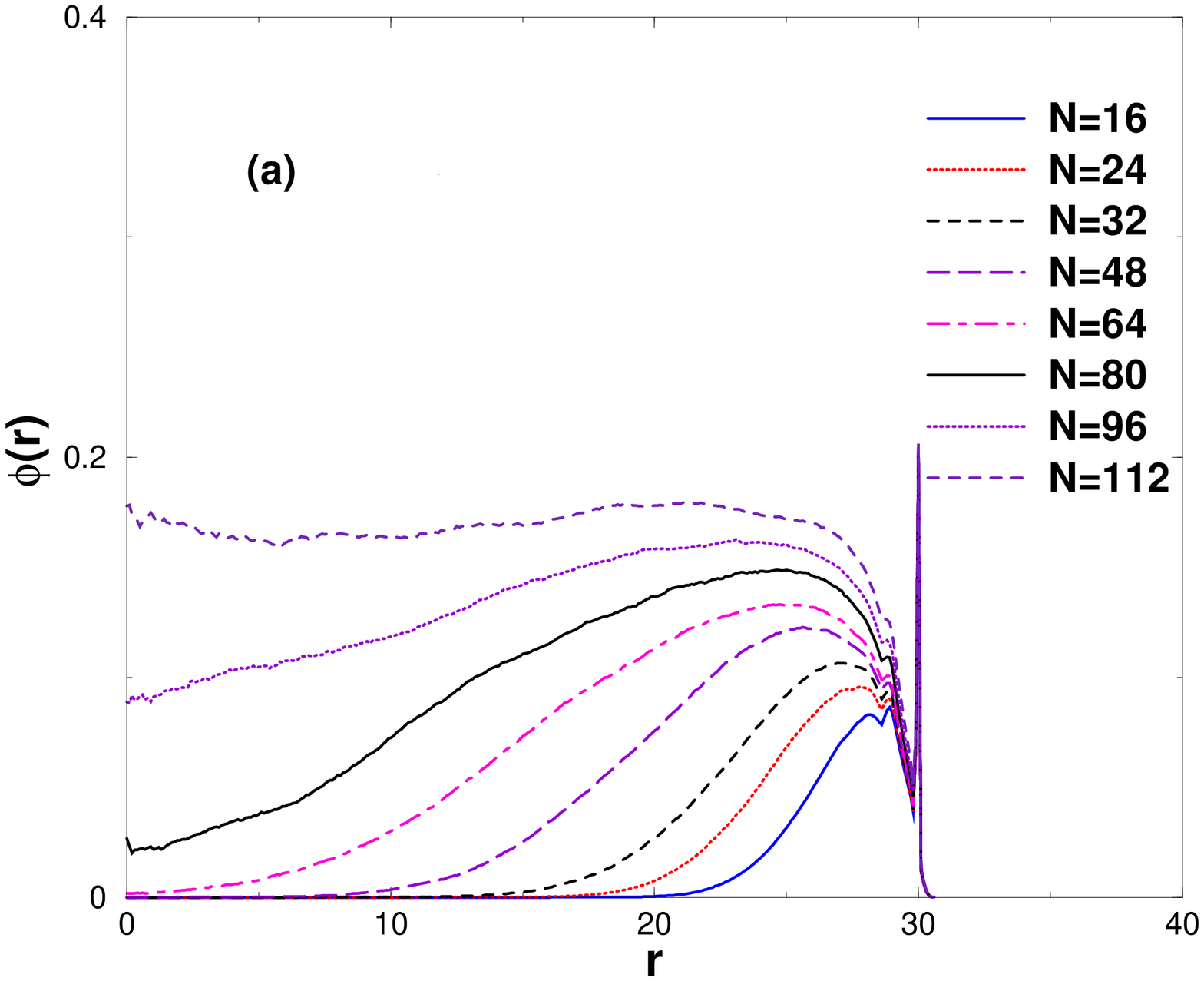}}
\centerline{\includegraphics[width=2.7in,
angle=-90]{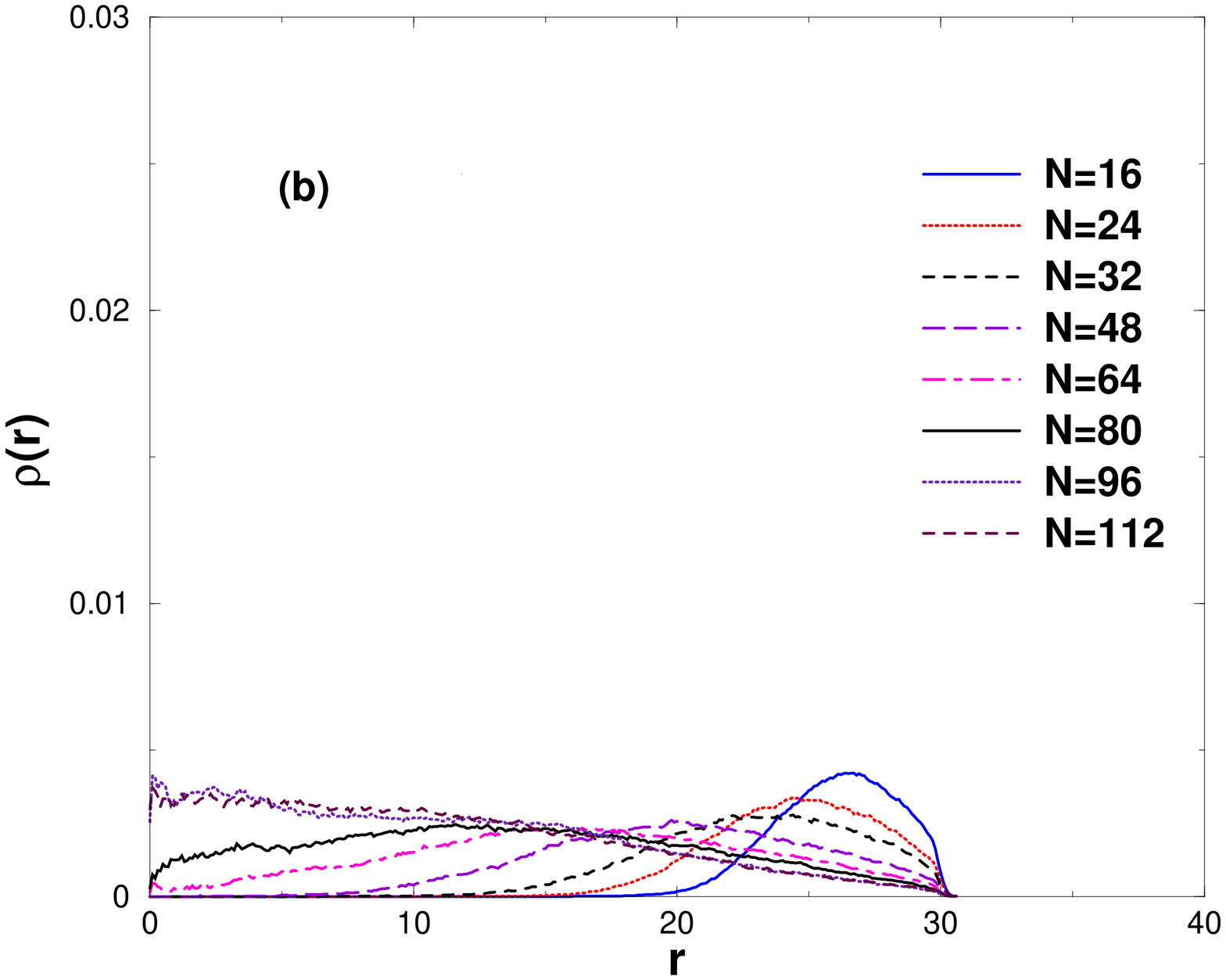}} \vspace*{8pt} \caption{Same as
Fig.~\ref{fig14}, but for R = 31, $\sigma = 0.022$
($N_{ch}=280$) \label{fig15}}
\end{figure}

\clearpage

\begin{figure}
\centerline{\includegraphics[width=2.7in,
angle=-90]{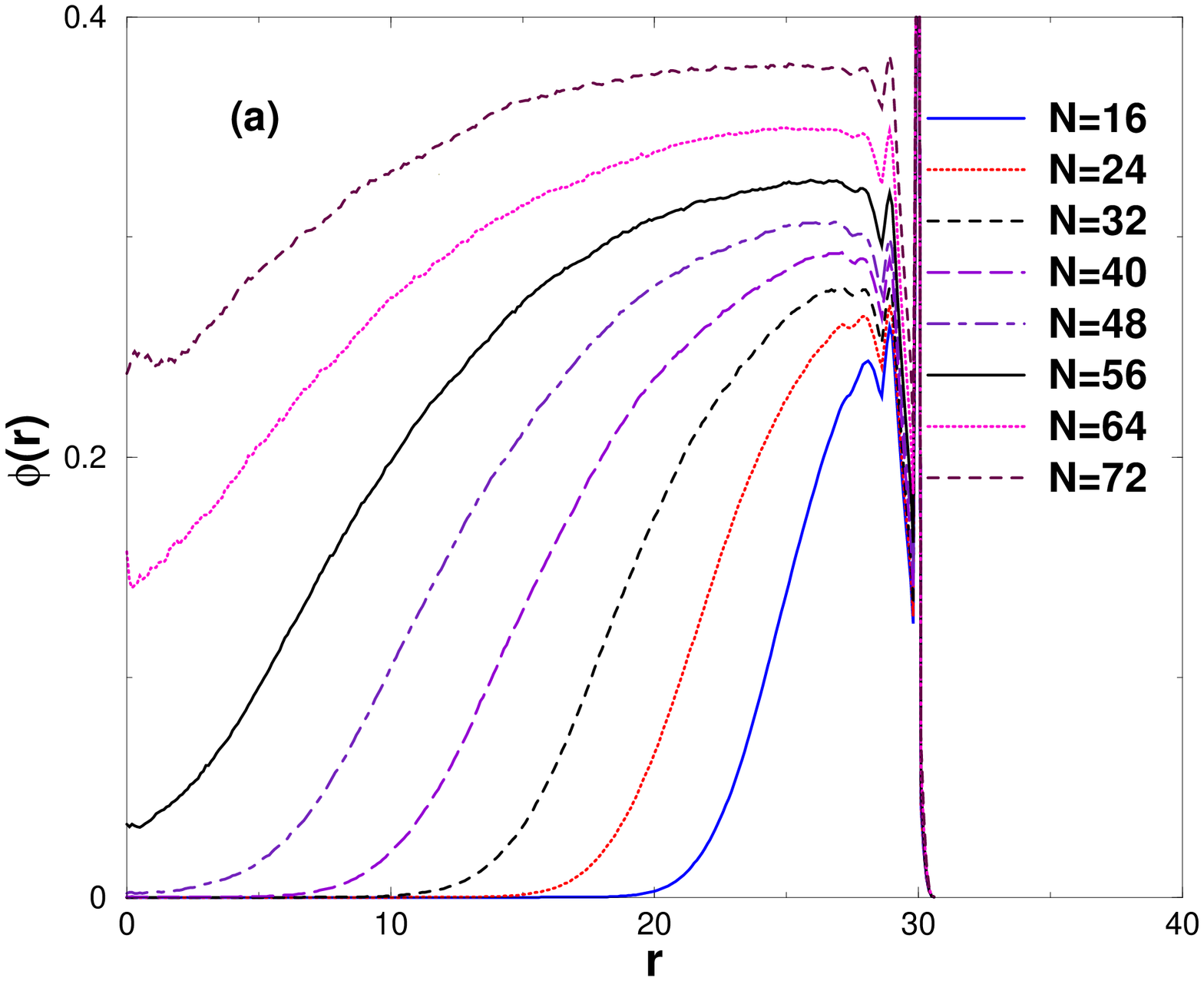}}
\centerline{\includegraphics[width=2.7in,
angle=-90]{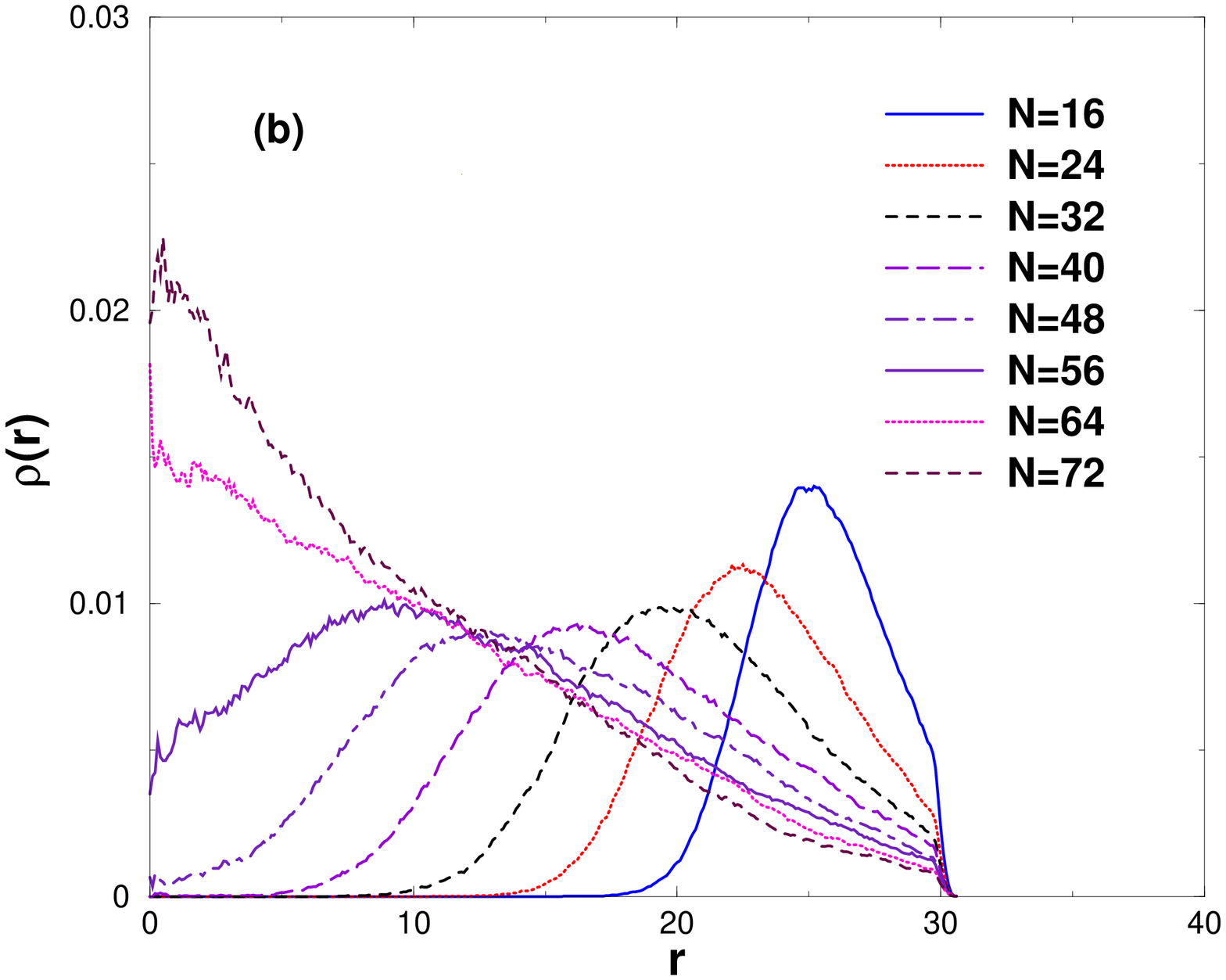}} \vspace*{8pt} \caption{Same as
Fig.~\ref{fig14}, but for R = 31, $\sigma = 0.0755$
($N_{ch}=960$) \label{fig16}}
\end{figure}

\clearpage

\begin{figure}
\centerline{\includegraphics[width=3.7in,
angle=-90]{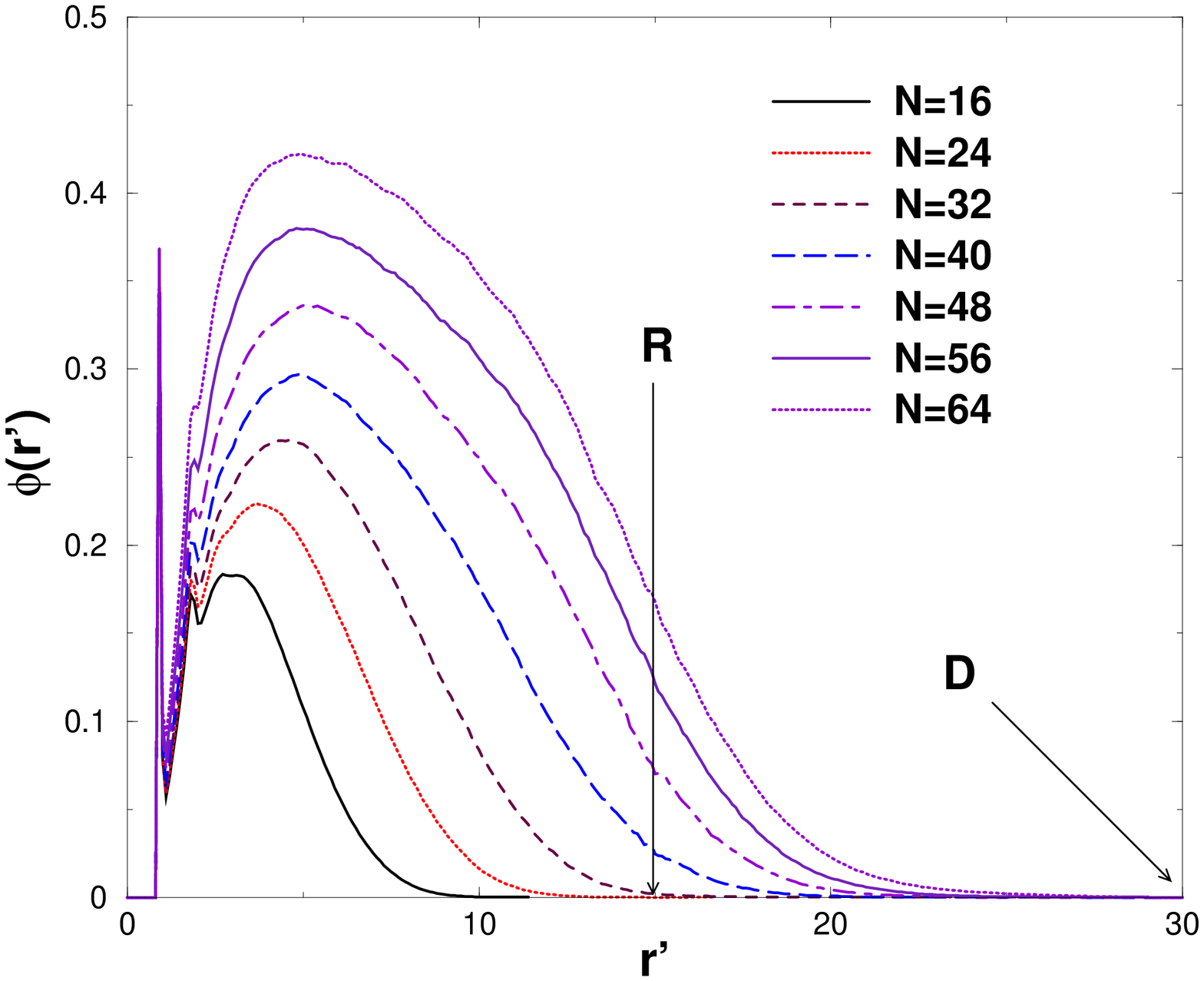}} \vspace*{8pt}
\centerline{\includegraphics[width=3.7in,
angle=-90]{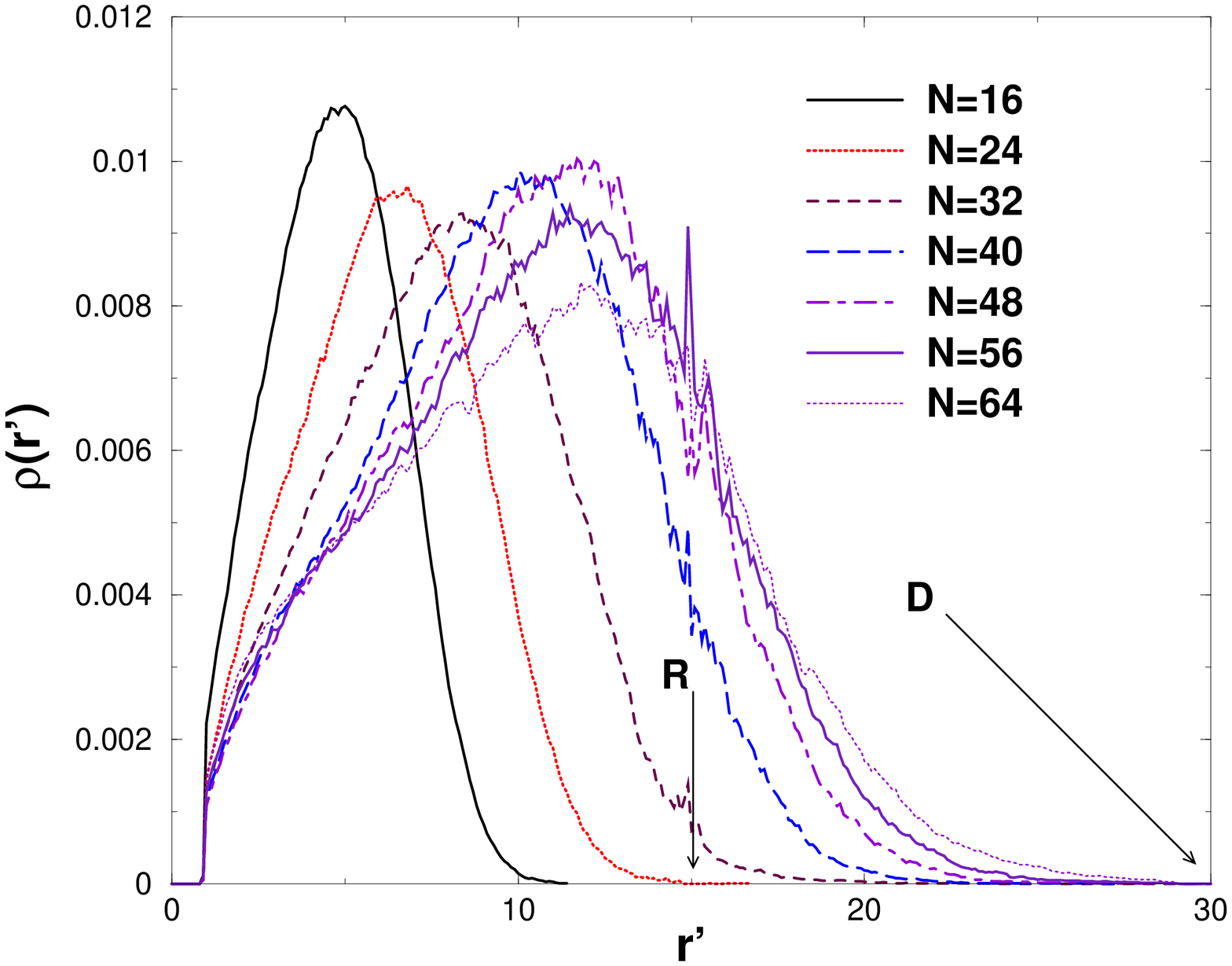}} \vspace*{8pt}
\caption{(a) Density profile $\phi (r')$ of the
monomers in the brush for $D=28$ at the grafting density $\sigma =
0.02$ and various chain lengths. Arrows point to the positions
of the cylinder axis and the maximum
distance from the grafting site that is possible in $z$-direction
inside the tube are marked by arrows. (b) Distribution $\rho (r')$
of the chain ends, for the same parameters as part
(a).\label{fig17}}
\end{figure}

\clearpage

\begin{figure}
\centerline{\includegraphics[width=2.7in,
angle=-90]{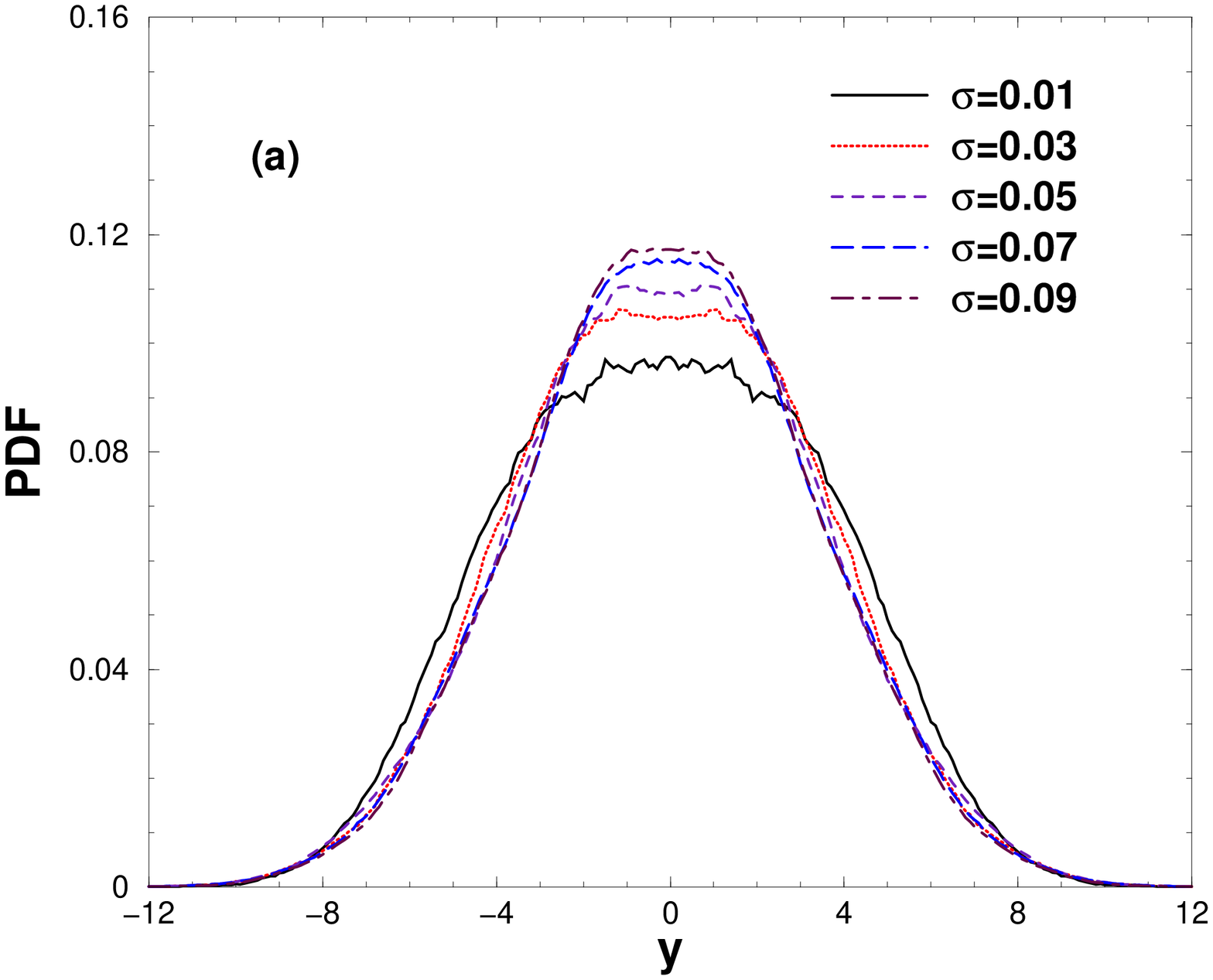}}
\centerline{\includegraphics[width=2.7in,
angle=-90]{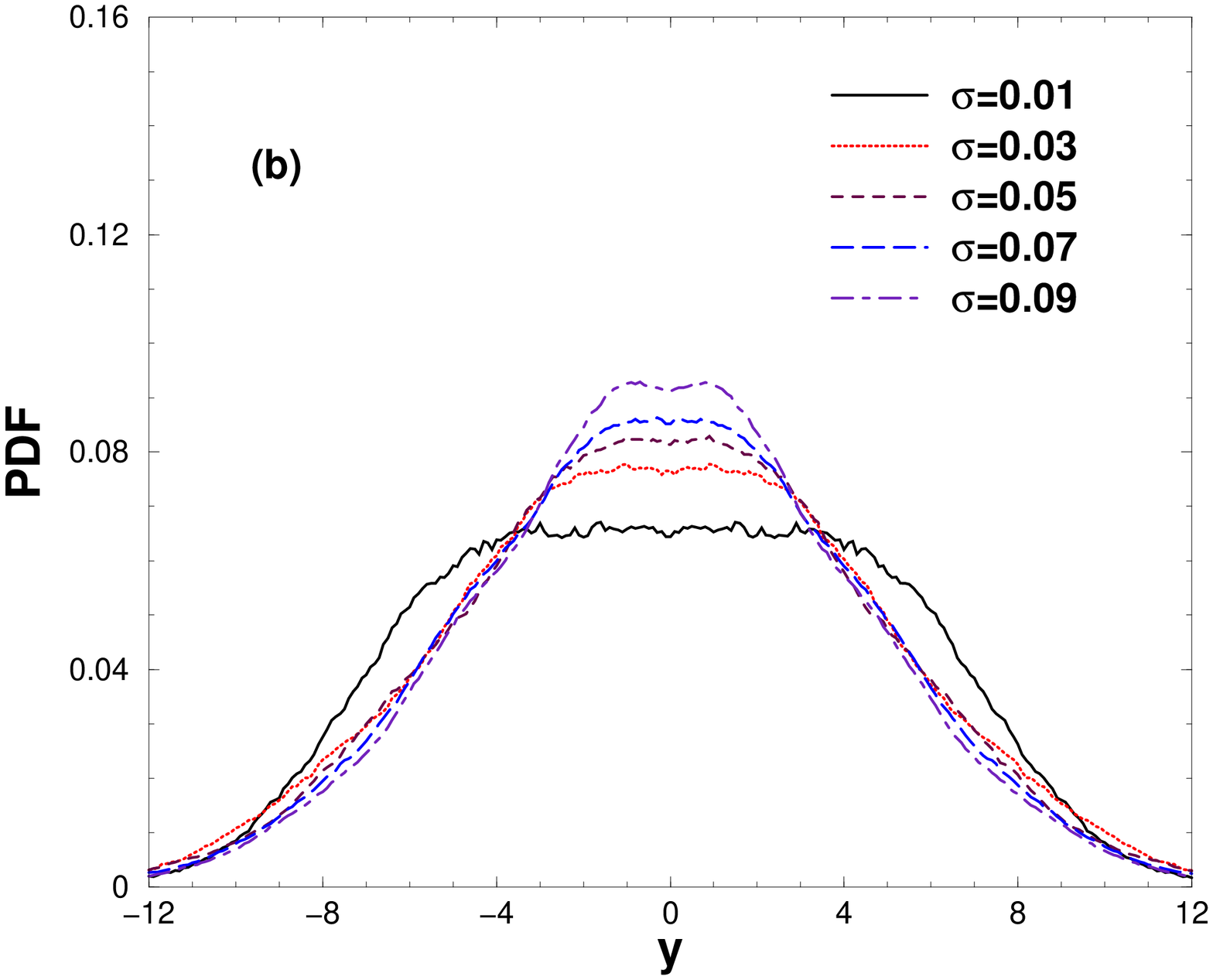}} \vspace*{8pt}
\caption{Probability distribution $\rho (y)$ of the
chain ends plotted vs. the $y$-coordinate along the tube axis, for
$D=30$ and two choices of $N,\; N=24$ (a) and $N=40$ (b). Several
values of the grafting density are included, as
indicated.\label{fig18}}
\end{figure}

\clearpage

\begin{figure}
\centerline{\includegraphics[width=3.7in,
angle=-90]{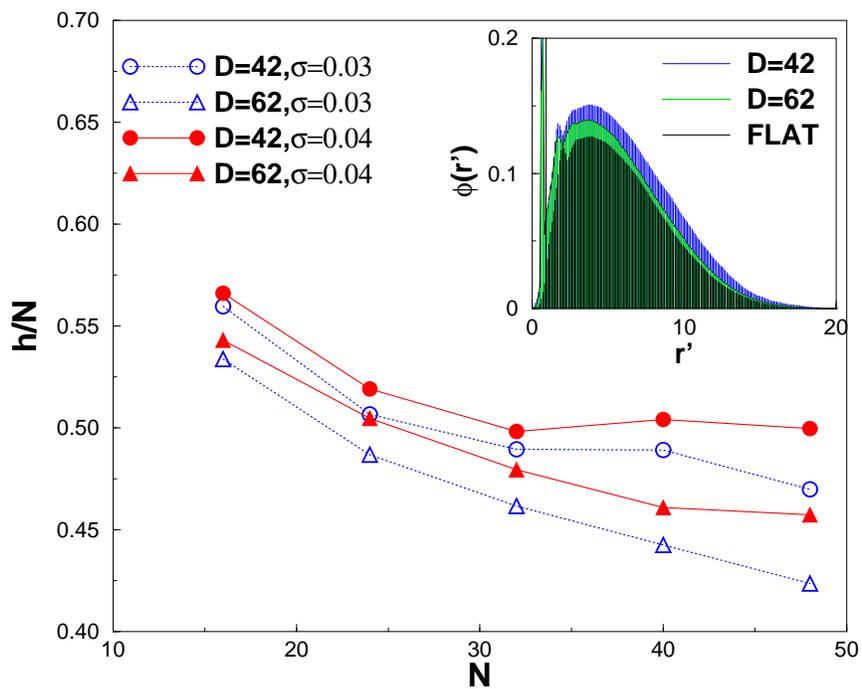}} \vspace*{8pt}
\caption{Plot of the reduced brush height $h/N$ 
in tubes with diameters $D=42$ and $62$ as
function of $N$ at two different grafting densities.
The inset shows the radial density distribution $\phi(r')$
for $N=32$ and $\sigma=0.03$ for a flat substrate (the lowest
$\phi(r')$), for a tube with $D=62$ (intermediate), and 
for a tube with $D=42$ (largest $\phi(r')$). \label{fig19}}
\end{figure}

\end{document}